\documentclass[sigconf]{acmart}
\usepackage{subcaption}
\usepackage{pifont}

\acmYear{2024}\copyrightyear{2024}
\setcopyright{rightsretained}
\acmConference[ASIA CCS '24]{ACM Asia Conference on Computer and Communications Security}{July 1--5, 2024}{Singapore, Singapore}
\acmBooktitle{ACM Asia Conference on Computer and Communications Security (ASIA CCS '24), July 1--5, 2024, Singapore, Singapore}
\acmDOI{10.1145/3634737.3645012}
\acmISBN{979-8-4007-0482-6/24/07}
\begin{document}
\title{Decoding the MITRE Engenuity ATT\&CK Enterprise Evaluation: An Analysis of EDR Performance in Real-World Environments}

\author{Xiangmin Shen}
\affiliation{%
  \institution{Northwestern University}
  \city{Evanston}
  \state{Illinois}
  \country{USA}}
\email{xiangminshen2019@u.northwestern.edu}

\author{Zhenyuan Li}
\affiliation{%
  \institution{Zhejiang University}
  \city{Hangzhou}
  \state{Zhejiang}
  \country{China}}
\email{lizhenyuan@zju.edu.cn}

\author{Graham Burleigh}
\affiliation{%
  \institution{Northwestern University}
  \city{Evanston}
  \state{Illinois}
  \country{USA}}
\email{grahamburleigh2022@u.northwestern.edu}

\author{Lingzhi Wang}
\affiliation{%
  \institution{Northwestern University}
  \city{Evanston}
  \state{Illinois}
  \country{USA}}
\email{lingzhiwang2025@u.northwestern.edu}

\author{Yan Chen}
\affiliation{%
  \institution{Northwestern University}
  \city{Evanston}
  \state{Illinois}
  \country{USA}}
\email{ychen@northwestern.edu}

\renewcommand{\shortauthors}{Shen, et al.}

\begin{abstract}
Endpoint detection and response (EDR) systems have emerged as a critical component of enterprise security solutions, effectively combating endpoint threats like APT attacks with extended lifecycles. In light of the growing significance of endpoint detection and response (EDR) systems, many cybersecurity providers have developed their own proprietary EDR solutions. It's crucial for users to assess the capabilities of these detection engines to make informed decisions about which products to choose. This is especially urgent given the market's size, which is expected to reach around 3.7 billion dollars by 2023 and is still expanding. MITRE is a leading organization in cyber threat analysis. In 2018, MITRE started to conduct annual APT emulations that cover major EDR vendors worldwide. Indicators include telemetry, detection and blocking capability, etc. Nevertheless, the evaluation results published by MITRE don't contain any further interpretations or suggestions. 

In this paper, we thoroughly analyzed MITRE evaluation results to gain further insights into real-world EDR systems under test. Specifically, we designed a whole-graph analysis method, which utilizes additional control flow and data flow information to measure the performance of EDR systems. Besides, we analyze MITRE evaluation's results over multiple years from various aspects, including detection coverage, detection confidence, detection modifier, data source, compatibility, etc.
Through the above studies, we have compiled a thorough summary of our findings and gained valuable insights from the evaluation results. 
We believe these summaries and insights can assist researchers, practitioners, and vendors in better understanding the strengths and limitations of mainstream EDR products.

\end{abstract}




% \settopmatter{printfolios=true}
% \settopmatter{printacmref=true}
\maketitle              % typeset the header of the contribution
\section{Introduction}
\label{sec:intro}
The digital revolution dramatically changed human life and brings new risks into daily life. 
Driven by profit, attackers in cyberspace have organized increasingly sophisticated attacks that affect many organizations and large corporations such as Siemens, Target, and Equifax.
These attacks resulted in millions of consumers' data being leaked \cite{barrett_how_nodate, staff_target:_2013, broad_israeli_2011} and other losses. 
Traditional network-based prevention and detection approaches can barely deal with these advanced attacks. 
Therefore, endpoint-based detection and response solutions (EDR) and extended solutions (XDR) receive extensive attention in academia and industry.
The market capitalization for the top 37 EDR companies has reached over 320 billion U.S. dollars~\cite{MarketCap} by 2022, the market size of 3.7 billion U.S. dollars~\cite{mordorintelligence_endpoint}. Meanwhile, both market capitalization and size are expected to grow fast.

As the number of homogeneous EDR products increases, it becomes increasingly difficult for users to choose the appropriate product.
Fair third-party evaluations with detailed interpretations of results are therefore necessary.
The challenges of conducting such evaluations are three-fold. 
Firstly, the evaluation methodology should be general enough to allow broad participation. The significance of the evaluation will be affected if its methodology only applies to a small set of security solutions.
Secondly, the evaluation should perform realistic and various attack emulations. The attack emulations in the laboratory setting are simple and have little variety. Sometimes, the attack information is even known before the evaluation, making it possible for the defense team to make ad-hoc configuration adjustments. Such attack emulations can not reveal the actual performance of endpoint security solutions against real-world threats.
Finally, the evaluation results should be reported with comprehensive interpretation. Without appropriate metrics and objective interpretation, the evaluation results are hard to understand, possibly leading to biased interpretation.

Although much effort has been made toward establishing a norm for security solutions, there is still no substantial benchmark in the security field. 
Several for-profit companies and organizations present their own evaluation results~\cite{noauthor_magic_nodate,noauthor_avcomparatives_nodate}. However, their methodologies are not transparent and could be biased for commercial reasons. 
In academia, several recent benchmark works~\cite{IDS, NIDS_dataset, ANIDS_dataset, IoT_benchmark, darpa_tc} focus on generating new or improving existing datasets. Although they are crucial initial steps, equally-important interpretations of evaluation results are still missing.
Several other benchmarking works \cite{SAST_benchmark,web_system_benchmark} attempt to expand the explainability of evaluation results. But their methodologies are specific to the target systems or platforms, making their evaluation methodologies and results not transferable. 

To standardize security evaluations, the MITRE Corporation has been conducting annual APT emulations to evaluate various security solutions based on its ATT\&CK framework \cite{noauthor_att&ck_nodate} since 2018. In each round of evaluation, MITRE will select one or more real-world APT groups and reconstruct their typical attack chains in controlled environments. 
The EDR products will be deployed in these environments in advance.
And the results will be collected and published by MITRE to summarize their performance. 
The results list the attack steps characterized by MITRE ATT\&CK techniques in attack emulations. For each EDR system, MITRE publishes its detection and protection performance on attack steps. The detection performance is described with MITRE-defined detection categories, indicating the amount of contextual information the EDR system provides with the alarm. The protection performance is illustrated by the step at which the attack is blocked.

While the datasets from MITRE's evaluation are valuable, the presentation of evaluation results has some apparent defects, preventing security practitioners from benefiting directly. 
These problems include missing whole-graph analysis, lacking comprehensive interpretation, and inconsistent evaluation framework. 
Concretely, MITRE only focuses on single-step detection results. However, the attacks that EDR systems fight against are sophisticated and involve multiple steps. EDR systems must consider the entire kill chain to provide satisfying detection and response services. Therefore, the whole-graph analysis capability is crucial in evaluating EDR systems.
Apart from conducting attack emulations and collecting results, interpreting the results is equally or not more critical to EDR system evaluation. However, MITRE makes minimal effort in interpreting the results. Lacking comprehensive interpretation prevents users from getting direct insights from the results and could lead to biased interpretations for vendors and customers.
MITRE has only conducted four evaluations so far. Understandably, its methodology has evolved, leading to several inconsistencies in the published results. Inconsistencies like these can be burdensome for users, forcing users to investigate the difference in every year's methodology.

To address these problems, we propose analysis methodologies on the MITRE evaluation dataset to perform fine-grained whole-graph analysis and holistic assessments of EDR systems' capabilities. 
Then, we apply our methodology to analyze MITRE evaluation datasets. 
We investigate all attack scenarios and construct causal relationship attack graphs to present causal relationships between attack steps. We evaluate EDR systems' attack reconstruction capability by conducting the connectivity analysis, examining whether the EDR system can reconstruct the complete attack kill chain. We also assess EDR systems' response capability via the effectiveness analysis. In the effectiveness analysis, we use protection performance as an indicator. Specifically, we examine at which step each EDR system responds to the attack and determine if the EDR system is effectively protecting the host. 
Moreover, we discuss the evaluation results from several practical perspectives to measure the detection and protection performance of individual techniques and EDR systems, including detection coverage, detection confidence, detection quality, data source, and compatibility. We also investigate the trend of performance change from these perspectives.

In summary, this paper makes the following contributions:

\vspace{-\topsep}
\begin{list}{\labelitemi}{\leftmargin=1.5em}
 \setlength{\topmargin}{0pt}
 \setlength{\itemsep}{0em}
 \setlength{\parskip}{0pt}
 \setlength{\parsep}{0pt}
    \item We design and implement new analysis methods to systemically interpret MITRE ATT\&CK evaluation's results, with evaluation dimensions including whole-graph analysis that explores the correlation capability of EDR systems and additional metrics to capture aspects of the evaluation results not covered by MITRE.

    \item We reconstruct several attack scenarios used in MITRE evaluation and apply whole-graph analysis to examine EDR systems' attack reconstruction and behavior correlation capabilities, which reveal whether an EDR system can effectively detect and respond to attacks.
     
    \item We propose a new evaluation metric and identify and highlight flaws in EDR systems. We also pinpoint a list of findings to shed light on areas that require improvement and offer suggestions to enhance the performance of EDR systems.
    
\end{list}

% sec 3,4,5各一个

% In Section \ref{sec:background}, we introduce the MITRE evaluation dataset and its problems. In Section \ref{sec:methodology}, we present our analysis methodologies for the MITRE evaluation dataset. In Section \ref{sec:measurement} and \ref{sec:trend}, we share our analysis results along with interpretations. Section \ref{sec:related} summarizes the related work, and Section \ref{sec:discussion} discusses additional problems that are not addressed in this paper. Finally, Section \ref{sec:conclusion} gives a conclusion, as well as future directions.

\section{Background}
\label{sec:background}
% We introduce the dataset provided by MITRE evaluations in~\S\ref{subsec:MITRELayout} and point out problems with the dataset in~\S\ref{subsec:background_problems}.

\subsection{MITRE ATT\&CK Evaluation}
\label{subsec:MITRELayout}

% \begin{figure*}
% \centering
% \includegraphics[width=0.9\textwidth]{figures/MITREworkflow.png}
% \vspace{-1em}
% \caption{MITRE Evaluation Methodology}
% \label{mitre_methodology}
% \vspace{-1em}
% \end{figure*}

MITRE ATT\&CK Evaluation is an APT emulation conducted yearly by MITRE Corporation, started in 2018. Its participants include most leading security companies, such as Palo Alto Networks, Fortinet, and CrowdStrike.
Each evaluation emulates attacks from well-known APT groups like APT3, APT29, and FIN7. Contrary to other attacks like malware and phishing, APT attacks are more complicated, involving multiple stages aiming for specific tasks. Together, those stages form a kill chain to achieve the final goals, such as stealing sensitive information or destroying valuable properties. The MITRE Corporation has established a set of Tactics, Techniques, and Procedures (TTPs)~\cite{noauthor_matrix_nodate} to outline each stage of the emulation process, which serve as a foundation for organizing steps in a kill chain. Tactics divide attack steps into 14 general stages, while techniques further distinguish attack steps according to the specific approach. In some cases, each technique can have associated sub-techniques, with additional details necessary to identify them accurately. Attacks performed in evaluations are illustrated step-wise, with individual steps associated with the techniques described above. Additionally, the information provided for each step in detection tests includes detection categories and modifiers, if applicable. The detection categories include the following types.

\begin{enumerate}
    \item \textit{Not Applicable:}
    The EDR system does not deploy a sensor on the given platform and thus has no visibility.
    \item \textit{None:}
    The EDR system deploys sensors on the given platform, but no data is available to show the event happened.
    \item \textit{Telemetry:}
    The EDR system knows the event happened but is unsure if they are malicious.
    \item \textit{General Behavior:}
    The EDR system knows the event happened and believes they are malicious. However, the system is unsure why and how the action was performed.
    \item \textit{Tactic:}
    The EDR system knows the event happened and believes they are malicious. The system knows why the action was performed but is unsure how the action was performed.
    \item \textit{Technique:}
    The EDR system knows the event happened and believes they are malicious. Additionally, the system knows why and how the action was performed.
\end{enumerate}

In addition to the detection categories, modifiers provide more context about the detection. The modifiers include \textit{delayed} and \textit{config change}.
\begin{enumerate}
    \item \textit{Delayed} means the alert appears significantly late compared to the time when the attack step happens.
    \item \textit{Config change} means the alert shows up due to ad-hoc configuration modifications. 
\end{enumerate}

During the latest two evaluations, a new scenario was introduced to test the protection ability of EDR systems. The results of each step in the protection tests are categorized into one of the following protection categories.

\begin{enumerate}
    \item \textit{Not Applicable:}
    The EDR system does not deploy a sensor on the given platform and thus has no visibility.
    \item \textit{None:}
    The EDR system deploys sensors on the given platform but does not block the malicious behavior.
    \item \textit{Blocked:}
    The EDR system successfully blocked the malicious behavior.
\end{enumerate}

To quantify the detection performance of EDR systems, MITRE defines four metrics to summarize each EDR system's capabilities at a high level: \textit{Visibility}, \textit{Telemetry Coverage}, \textit{Analytic Coverage} and \textit{Detection Count}. 
\begin{enumerate}
    \item \textit{Telemetry Coverage} is the number of detected steps with the telemetry level detection. This is the minimum requirement for a step to be visible, as telemetry detection only confirms an event has happened but wouldn't trigger an alarm.
    \item \textit{Analytic Coverage} is the number of detected steps with some contextual information like the intention and the approach taken. Since only detection above the telemetry level is reported as malicious behavior, the analytic coverage reflects a system's ability to detect threats from the available data.
    \item \textit{Visibility} is the number of steps with at least a telemetry detection. Note that this metric counts the number of steps in the union of \textit{Telemetry Coverage} and \textit{Analytic Coverage}.
    \item \textit{Detection Count} is the total number of detection made in the attack campaign. This number could be larger than the total steps, as multiple detections in different categories might be reported at a certain step.
\end{enumerate}

% Figure \ref{mitre_results} shows an example of the MITRE evaluation results SentinelOne received.
% \begin{figure*}
% \centering
% \includegraphics[width=\linewidth]{figures/MITRE_results.pdf}
% \vspace{-1em}
% \caption{MITRE Evaluation Results}
% \label{mitre_results}
% \vspace{-1em}
% \end{figure*}

\subsection{Limitations of ATT\&CK Evaluation}
\label{subsec:background_problems}
The MITRE evaluation has made significant contributions to establishing an evaluation standard for EDR solutions. However, many limitations still need to be addressed and improved upon.

\subsubsection{Missing whole-graph analysis}
The security field has been shifting from single-point detection to graph-based detection. The single-point detection can only detect a single step in an attack without providing an overview of the entire attack pattern. In contrast, whole graph-based detection utilizes contextual information to construct a comprehensive graph that depicts behavior and searches for threats. For modern endpoint APT defense, relying solely on single-point detection is inadequate for two reasons: Firstly, single-point detection is vulnerable to complex and sophisticated attacks that can evade traditional detection methods. Attackers can use multiple techniques to bypass single-point detection. Secondly, single-point detection focuses only on one aspect without considering contextual information, such as control and data flow, which limits perspective and could lead to false positive alarms. 

Provenance graph-based detection~\cite{provenance, backtracking_intrusion} overcomes these shortcomings by taking additional contextual information into account and obtaining a comprehensive view of the endpoint. Even if a single step is not identified as malicious from a single-point perspective, it can still be determined as part of the kill chain through control flow and data flow connections with other malicious behaviors. Moreover, using such correlations, malicious behaviors can be better distinguished from benign activities, reducing the number of false alarms. Most state-of-the-art endpoint detection work incorporates provenance graphs and their derivatives as part of their framework.

\begin{figure}
\centering
\includegraphics[width=\linewidth]{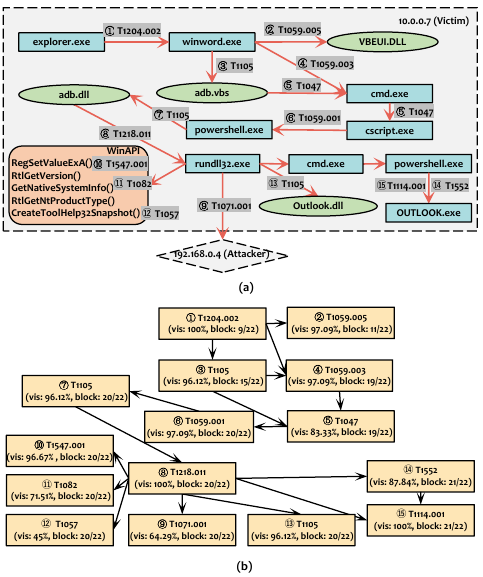}
\vspace{-2em}
\caption{The attack graphs for scenario 1 in Wizard Spider+Sandworm (2022) evaluation. (a) The actual attack graph. The nodes are system entities like processes and files. The edges represent system events characterized by MITRE ATT\&CK techniques IDs. The numbers denote the order of events. (b) The causal relationship attack graph. The nodes are attack steps characterized by MITRE ATT\&CK techniques IDs. The edges represent causal relationships between attack steps. The nodes also contain the visibility of their corresponding techniques among all EDR systems and the number of EDR systems that blocked this attack before and at this step.}
\label{wizard-spider-attack-2-1}
\vspace{-1em}
\end{figure}

As discussed in the \S\ref{subsec:MITRELayout}, MITRE uses a sequence of techniques to describe attack scenarios. However, the execution of kill chains in attack emulations hardly follows a linear pattern.
Although such sequential representation emphasizes the detection performance on single steps, it obscures the causal and spatial aspects of the attack scenario.
Fig. \ref{wizard-spider-attack-2-1} shows an example of the attack graphs we constructed from an attack emulated in Wizard Spider+Sandworm (2022) evaluation. Without the graph, it's hard to understand the importance of each step in event correlation. For instance, \texttt{rundll32.exe} loads the downloaded malicious DLL file \texttt{adb.dll} to perform the following attack steps at step \ding{179}. Missing this step in the scope makes it hard to correlate the following attack steps with the previous setup steps. However, missing other less important steps, like \texttt{winword.exe} loading a malicious DLL file \texttt{VBDUI.DLL} at step \ding{173} does not affect the connectivity nor the causal relationship. With the help of the graph, we can investigate 1) whether the EDR systems can detect all crucial steps and 2) whether they can correlate events along the attack chain to reconstruct the attack chain and protect the system.

\subsubsection{Lacking comprehensive interpretation}
Another crucial part missing in MITRE evaluations is comprehensive interpretations of the results. Multiple companies claim they have achieved perfect or near-perfect performance in these evaluations. However, such claims contradict the results presented on the MITRE website. CrowdStrike~\cite{Crowdstrike} claims to have received 100\%  detection coverage across all 20 steps of the Carbanak+Fin7 evaluation. However, such a claim does not align with the evaluation results. Specifically, the `20 steps' are comprised of 174 substeps, where CrowdStrike failed to catch 22 out of 174 substeps and failed to generate alarms for 88 out of 152 visible steps. Other vendors have exhibited similar results.

Lacking clear interpretation leave room for EDR vendors to twerk the results, ironically going against the original intention of MITRE evaluation. Thus, it is necessary to add a comprehensive and objective interpretation on top of the MITRE Evaluation raw results.

% edge:T1111(DR: 70%, block: 3/19)

\subsubsection{Inconsistent evaluation framework}
The MITRE Engenuity started the evaluation project in 2018. Since then, the evaluation approaches and terminology have been changing yearly, making it hard to compare the detection performance from different evaluations.
For example, in the first APT3 evaluation, there are six detection categories, including \textit{None}, \textit{Telemetry}, \textit{Indicator of Compromise}, \textit{Enrichment}, \textit{General Behavior}, and \textit{Specific Behavior}. In the most recent Wizard Spider+Sandworm (2022) evaluation, there are five detection categories, including \textit{None}, \textit{Telemetry}, \textit{General}, \textit{Tactic}, and \textit{Technique}. Although \textit{None} and \textit{Telemetry} remain the same, MITRE didn't map the rest of the detection categories. Besides, MITRE has used several versions of detection modifiers and even changed the definition of performance metrics over the years. In the most recent Wizard Spider+Sandworm (2022) evaluation, MITRE changed how detection numbers are counted. Instead of counting multiple detections on a single step, MITRE only recorded the detection with the most contextual information. In this way, visibility is the sum of telemetry and analytic coverage. The detection count metric is deleted since it is always the same as visibility.

% \subsection{Our Goal}
Thus, in this paper, to bridge the gap between direct results and insight, we aim to provide a comprehensive interpretation of MITRE Engenuity Evaluation results. We also try to extract the consistent aspects from the different terminology used in each year's evaluation to establish a compatible interpretation framework to compare evaluation results from different years.

\section{Interpretation Methodology}
\label{sec:methodology}
\subsection{Overview}
\begin{figure}
\centering
\includegraphics[width=\linewidth]{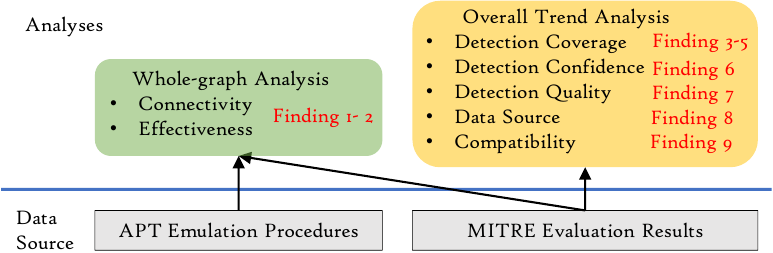}
\vspace{-1em}
\caption{An overview of our analysis methodologies.}
\label{analysis-overview}
\vspace{-1em}
\end{figure}

We start this section by introducing the dataset in \S\ref{subsec:dataset}. Then, we elucidate two approaches to analyze this dataset: (1) whole-graph analysis and (2) overall statistical and trend analysis.  
Fig. \ref{analysis-overview} present an overview of our analyses. 
In the whole-graph analysis, we studied techniques provided in the evaluation results and APT emulation procedures published by MITRE Center for Threat-Informed Defense~\cite{adversary_emulation_library} to construct several causal relationship attack graphs. Via connectivity analysis and effectiveness analysis of the causal relationship attack graphs, we investigate the EDR systems' attack graph-level correlation and reconstruction capabilities.
In the overall trend analysis, we investigate the detection performance of EDR systems on various techniques through several perspectives over the years. We aim to provide insights into the strengths and areas requiring enhancement within intrusion detection.
We present the detailed methodologies in \S\ref{sebsec:whole-graph} and \S\ref{subsec:overall-trend}, respectively. 
% By adopting this dual-pronged approach, we aim to provide a comprehensive analysis and derive insightful observations from the evaluation results.

\begin{table*}[]
\vspace{-0.5em}
\centering
\caption{MITRE Engenuity Dataset Summary} 
\vspace{-0.5em}
\begin{tabular}{|l|c|c|c|c|}
\hline
Round of Evaluation      & Participants & Steps & Techniques & \# of Detection Made \\ \hline\hline
APT3 (2018)              &      12      &   136    &     51     & 1970   \\
\hline
APT29 (2019)             &      21      &   134    &     53     & 3982   \\
\hline
Carbanak+FIN7 (2020)     &      29      &   174    &     46     & 7350   \\
\hline
Wizard Spider+Sandworm (2022) &      30      &   109    &     46  & 3098 \\
\hline\hline
Total                    &      37      &   N/A    &     82     & 16.4k    \\
\hline
\end{tabular}
\vspace{-0.5em}
\label{MITREDatasetSummary}
\end{table*}

\subsection{Dataset}
\label{subsec:dataset}
The MITRE Engenuity has conducted four evaluations so far. Each evaluation selects one or two Advanced Persistent Threat (APT) groups and emulates several attack scenarios using the APT groups' toolkits. In total, attack scenarios consist of hundreds of steps involving dozens of techniques. For every attack scenario, MITRE Engenuity will conduct a detection test and a protection test. In detection tests, the attack kill chain will be executed without intervention. MITRE will assess whether every attack step is detected and the extent of contextual information provided during detection. While in protection tests, the defense team can intervene and stop the consequential steps in the kill chain. In this way, MITRE can assess whether an attack is contained or blocked and at which step. Therefore, the published dataset presents the results in a step-wise manner. For the detection results, data entries specify the following information about an attack step: (1) the MITRE ATT\&CK technique that corresponds to the step, (2) whether the step is detected and how much contextual information is provided by the EDR system, (3) the data sources associated with detection, and (4) other miscellaneous information. The protection results also contain the MITRE ATT\&CK technique corresponding to the step. Besides, instead of the detection-related information, the results show what kind of protection is triggered at each step. The definition of detection and protection categories is detailed in \S\ref{subsec:MITRELayout}. Table \ref{MITREDatasetSummary} outlines detailed information about the dataset associated with each campaign. Overall, we analyzed 16.4k detection results from all vendors in all published evaluation results.

\subsection{Whole-graph Analysis}
\label{sebsec:whole-graph}

\subsubsection{Attack Graph Construction}
Due to the importance of provenance graph-based detection capabilities, we create a causal relationship graph model for each scenario to replace the sequential layout of MITRE evaluation results.
A causal relationship attack graph is a directed graph in which the node represents an attack step in the kill chain, and the edge denotes the causal relationship between two attack steps.
Constructing a causal relationship attack graph model involves nodes construction and edges construction. 

Since MITRE only describes each step in terms of techniques, our initial step is to thoroughly examine each step's corresponding procedure, which is subsequently classified into descriptive and causal categories. The descriptive techniques solely describe specific features of a step without establishing any causal connections with other entities (e.g., \textit{Encrypted Channel}). In contrast, the causal techniques interact with other entities, such as creating a process or writing to a file, thereby establishing causal relationships with other steps (e.g., \textit{Ingress Tool Transfer}). All steps corresponding to causal techniques become the nodes in our causal relationship attack graph. 

After classifying each step, we examine the subject and object of each step to establish causal relationships. When we connect the subject to the object, the edges are constructed in the graph. We generally consider two kinds of causal relationships: control flow and data flow. 
Firstly, the control flow creates causal relationships via process creation. If an attack step is performed by a process created in a previous step, then the two steps establish a control flow causal relationship.
Secondly, the data flow creates causal relationships via communication over files. If an attack step reads a file written in a previous step, the two steps establish a data flow causal relationship.
Fig.~\ref{wizard-spider-attack-2-1}(b) and \ref{wizard-spider-attack-2-2}(b) in the Appendix show two examples of causal relationship attack graphs.
% Fig. \ref{fig:causal_attack_graph} shows an overview of the causal relationship attack graphs constructed from attack scenarios in the Wizard+Sandworm evaluation.

% \begin{figure}[!h]
%      \centering
%      \begin{subfigure}[b]{0.235\textwidth}
%         \centering
%         \includegraphics[width=\linewidth]{figures/wizard_spider-sandworm/wizard_spiderCasualGraph.pdf}  
%         \caption{Scenario 1: Wizard Spider.}
%         \label{wizardspidercasualgraph}
%         \vspace{-.5em}
%      \end{subfigure}
%      \hfill
%      \begin{subfigure}[b]{0.235\textwidth}
%         \centering
%         \includegraphics[width=\linewidth]{figures/wizard_spider-sandworm/sandwormCasualGraph.pdf}  
%         \caption{Scenario 2: Sandworm.}
%         \label{sandwormcasualgraph}
%         \vspace{-.5em}
%      \end{subfigure}
%         \caption{The casual relationship attack graphs from two scenarios in the Wizard Spider+Sandworm evaluation. Each scenario involves three hosts. The entry points on each host are marked with red at the top.}
%         \label{fig:causal_attack_graph}
%     \vspace{-.5em}
% \end{figure}

\subsubsection{Attack Graph Analysis}
After constructing a causal relationship attack graph, we analyze EDR systems' detection and protection performance from the attack graph perspective. We aim to answer the following two questions: (1) whether the EDR systems can fully reconstruct the attack kill chain; (2) whether the EDR systems can effectively aggregate behaviors along the kill chain to understand its severity. 
We answer the first question by conducting a connectivity analysis on the causal relationship attack graphs constructed from the detection results.
As for the second question, we analyze the protection performance to examine the EDR systems' effectiveness. 
An effective EDR system should detect and respond to threats at the appropriate time. We study several attack cases and investigate when EDR systems block the kill chain.
% When MITRE presented the evaluation results, they only focused on the detection performance without punishing false positive alarms. Intuitively, we address this problem by incorporating protection results since an EDR system is effective only if it detects and responds to a threat.
% Specifically, when we plug in the detection results, the casual relationship attack graphs allow us to examine the completeness of the kill chain detection, which is missing from the origin result layout. Additionally, we plug in detection and protection results and put them side by side to determine whether the EDR systems used their attack graph effectively.

\noindent\textbf{Connectivity Analysis}:
% \textcolor{red}{aiming at attack chain reconstruction capability.} 
We examine the kill chain visibility by counting the connected components in the causal relationship attack graph and comparing them with the ground truth. 
Suppose a campaign involves attacks on three individual hosts, leading to three separate kill chains among those hosts. If the number of connected components on the graph is more than three, one or more kill chains are broken into multiple small segments. If the number is less than three, at least one kill chain is completely missing in the detection. If the number equals three, we check if the three segments match the ground truth. In this way, we use the connected components as a metric to evaluate EDR systems' attack reconstruction capability.

%Note that if leaves are missing from the attack graph, it won't lead to segmentation and thus won't affect the completeness of the kill chain detection. 

\noindent\textbf{Effectiveness Analysis}:
We examine which step a blockage is triggered given each vendor's detection results to analyze how effectively the EDR systems use the graph information. 
We assume EDR systems are knowledgeable about the maliciousness of different behaviors, and they would block the attack once the severity of existing behaviors accumulates to a certain threshold. 
We select a few attacks to perform case studies. For each case, we manually determine a step on the attack kill chain when the malicious intention is evident as the baseline. Then, for each EDR system, we compare at which step the attack is blocked with the baseline. 
If the attack is blocked earlier than the baseline, it suggests the EDR system adopts an aggressive strategy in defense response. In this case, benign behaviors could be incorrectly classified as malicious, leading to unpredictable problems.
If the attack is blocked later than the baseline, it suggests the EDR system cannot react to threats in time. Such delay could allow the attacks to happen unhindered.
% We especially want to see whether the blockage is triggered once the indicator of compromise appears or if the EDR system waits until later to block it. 
% Besides, we also analyze how the detection category influences blockage decisions. We wonder how detection with different contextual information would impact the protection. 
% \textcolor{red}{How to calculate the severity score? 1. closer to the block has higher weight; 2. higher frequency has higher weight; 3. consider single and combination}

\subsection{Overall Trend Analysis}
\label{subsec:overall-trend}
Besides analyzing the evaluation results on the attack graph, we also investigate the detection and protection performance of individual techniques and EDR systems over the years. 
% Specifically, we are looking to explore some numbers to represent the overall performance of the entire industry and exam their changes over the years.

\subsubsection{Detection Coverage}
MITRE presents the evaluation from the vendors' perspective. Each vendor's performance is reflected by its analytic coverage, telemetry coverage, and visibility on all attack steps. We want to investigate EDR systems' performance from another perspective: how well all vendors can detect each technique. We analyze the detection coverage of a technique from two perspectives: visibility and analytic coverage. Like MITRE's metrics, we determine the visibility of a technique by calculating the ratio of EDR systems aware of the corresponding behaviors. We determine the analytic coverage of a technique by calculating the percentage of the vendors that successfully detected the corresponding step. 
% It is hard to detect if a step is detected only by a few vendors. Formally, the visibility and analytic coverage can be calculated as follows.
\begin{enumerate}
    \item \textit{Visibility}  
    \[
    V(x) = \frac{S_{v}(x)}{S_{t}(x)}
    \]
    Where $x$ is a specific EDR system or a specific technique. $V(x)$ is the visibility score of the given $x$. $S_{v}(x)$ is the number of visible substeps of the given $x$, and $S_{t}(x)$ is the total substeps related to the given $x$. 
    \item \textit{Analytic Coverage}
    \[
    A(x) = \frac{S_{a}(x)}{S_{v}(x)}
    \]
    Where $A(x)$ is the analytics score of the given $x$, $S_{a}(x)$ is the number of detections beyond the telemetry level of the given $x$, and $S_{v}(x)$ is the number of visible substeps of the given $x$.
\end{enumerate}
% \textcolor{red}{use a score calculation example to illustrate how it is calculated.}
% We want to give a list of easy-to-detect and hard-to-detect steps. 

\subsubsection{Detection Confidence}
Although the three metrics adopted from MITRE evaluation provide helpful information, some aspects are missing. Specifically, those metrics do not take different detection categories into account. In other words, a system reporting all malicious behaviors in the general behavior level would receive identical scores as a system reporting all malicious behaviors in the technique level. We propose an additional metric: \textit{confidence} to address this issue. 

Confidence is a weighted score calculated by multiplying the percentage of detection made from different detection methods by the corresponding weight multiplier. A high confidence score suggests more details about the malicious behavior are provided.
\[
C(x) = \frac{4D_{te}(x) + 3D_{ta}(x) + 2D_{ge}(x) + D_{tel}(x)}{4D_{v}(x)}
\]
Where $C(x)$ is the confidence score of a given technique or EDR system $x$. $D_{te}(x)$ is the number of technique detection of the given $x$, $D_{ta}$ the number of tactic detection of the given $x$, $D_{ge}$ the number of general behavior detection of the given $x$, $D_{tel}$ the number of telemetry detection of the given $x$, and $D_{v}$ the total visible substeps of the given $x$. The multiplier associated with each variable indicates the granularity of the detection results, with four being the most detailed and one being the most general. Since MITRE evaluation provides four levels of granularity for detection results, we intuitively use 1 through 4 as the multipliers. They can be further adjusted if more detailed data is available.

% \textcolor{red}{use a score calculation example to illustrate how it is calculated.}

\subsubsection{Detection Quality}
Another aspect missing from MITRE-provided metrics is the negative modifiers. A system reporting all alarms with significant delay or configuration change receives identical scores as a system reporting all alarms with no delay and no configuration change.
To examine the presence of modifiers quantitatively, we propose a \textit{quality} metric for techniques and EDR systems. For a technique, the quality score is the ratio of visible substeps without negative modifiers to the total visible substeps. A high-quality score implies low detection latency and adequate out-of-box usability. 
\[
Q(x) = \frac{S_{m}(x)}{S_{v}(x)}
\]
Where $S_{m}(x)$ is the number of visible substeps without negative modifiers of a given technique or EDR system $x$, and $S_{v}(x)$ is the total visible substeps of the given $x$. We treat all the negative modifiers equally since they are all related to manual adjustments or analyses.

% \textcolor{red}{use a score calculation example to illustrate how it is calculated.}
% We observe which technique is associated with each modifier most frequently.

\subsubsection{Data Source}
Besides the quantitative analysis specified above, we investigate the data sources used in evaluations via a rather qualitative approach. Specifically, we compare the data sources used in each year's evaluation to examine the scope of data sources used in EDR systems. We also discuss the frequency of a data source used in each evaluation to investigate the importance of the data source.
% We tally the data sources used to detect each technique and investigate the data source most benefits a specific technique.

\subsubsection{Compatibility}
We investigate EDR systems' compatibility from two perspectives: availability and performance. We examine the availability by calculating the ratio of EDR systems that support a given platform. Since MITRE Engenuity evaluations only involved Windows and Linux platforms so far, we will focus on the availability of these two platforms. Besides, we compare the detection performance on different platforms.

\section{Whole-graph Analysis}
\label{sec:measurement}

Since whole-graph analysis is specific to the attack scenarios, we use attack scenarios in the Wizard Spider+Sandworm (2022) evaluation as the cases to perform our whole-graph analysis.
We constructed casual relationship attack graphs at the procedure level for all attack scenarios in Wizard Spider+Sandworm Evaluation. Fig. \ref{wizard-spider-attack-2-1} and \ref{wizard-spider-attack-2-2} in the Appendix present two examples of constructing the causal relationship attack graphs. 
% We first build an attack graph comprised of system entities, like processes and files, and their interactions denoted by MITRE technique IDs as shown in Fig. \ref{wizard-spider-attack-2-1}(a) and \ref{wizard-spider-attack-2-2}(a). Then, we extract the causal relationship between interactions and construct the causal relationship attack graph as shown in Fig. \ref{wizard-spider-attack-2-1}(b) and \ref{wizard-spider-attack-2-2}(b).

\subsection{Connectivity Analysis} We analyze the attack graph connectivity by calculating the number of connected components in the casual relationship attack graph generated from the visible steps of each vendor and comparing it with the ground truth.
% Fig. \ref{freqsegbargraph} shows the distribution of segmentation counts among vendors from Wizard Spider+Sandworm (2022) Evaluation. 
There are six hosts involved throughout the attack emulation. Thus, there should be six segments. One of the six hosts runs under the Linux environment, and the other five are under the Windows environment. 22 vendors support Linux environment data collection and detection out of 30 participants. Therefore, we divide the vendors into two groups according to their Linux platform compatibility. Of the 22 vendors that support the Linux platform, three have more than six segments (Rapid7 has 13, Cisco has 10, and Cylance has 11). Of the eight vendors that don't support the Linux platform, two have more than five segments (Deep Instinct has nine, and ReaQta has six). 25 out of 30 (83.3\%) participants can obtain a visibly connected subgraph containing all attack steps. Thus, we conclude that most vendors can see the connection between attack steps on a graph level.

% \begin{figure}[]
% %\hspace*{.5cm}
% \centering
% \includegraphics[width=0.96\linewidth]{figures/wizard_spider-sandworm/freq_vs_seg.png} 
% \vspace{-1.5em}
% \caption{Distribution of segmentation counts among vendors. The x-axis shows the number of segmentation counts, while the y-axis displays the number of vendors with that many segmentations.}
% \label{freqsegbargraph}
% \vspace{-1.5em}
% \end{figure}

\subsection{Effectiveness Analysis} 
\label{subsec:effectiveness}
We analyze all protection evaluation scenarios and discuss two as case studies to see how effectively the EDR systems use the graph information. Table \ref{protectiontesttable} summarizes the results of nine protection tests. 22 vendors participated in the protection tests. Test 7 is conducted on Linux. Since five participants didn't support the Linux platform, only 17 were in Test 7. Fig. \ref{wizard-spider-attack-2-1} and \ref{wizard-spider-attack-2-2} in the appendix presents the two cases we will discuss in detail.

\begin{table}[]
\centering
\vspace{-0.5em}
\caption{Summary of Protection Test Results}
\vspace{-0.5em}
\label{protectiontesttable}
\begin{tabular}{|c|c|c|c|}
\hline
Test & \# of Blockage & \# of Participants & Protection Rate \\ \hline
1    & 21             & 22                 & 95.5\%          \\ \hline
2    & 21             & 22                 & 95.5\%          \\ \hline
3    & 16             & 22                 & 72.7\%          \\ \hline
4    & 12             & 22                 & 54.5\%          \\ \hline
5    & 15             & 22                 & 68.2\%          \\ \hline
6    & 20             & 22                 & 90.9\%          \\ \hline
7    & 9              & 17                 & 52.9\%          \\ \hline
8    & 20             & 22                 & 90.9\%          \\ \hline
9    & 18             & 22                 & 81.8\%          \\ \hline
\end{tabular}
\vspace{-1.5em}
\end{table}

\noindent\textbf{Scenario 1: Emotet Initial Compromise, Persistence, and Collection.}

Fig.~\ref{wizard-spider-attack-2-1} shows a detailed attack graph of this scenario. In this scenario, the adversary sent a Word document over email, which contained obfuscated VBA macros that downloaded and executed a malicious DLL based on the malware Emotet. The malicious DLL then established a command and control (C\&C) session with the adversary server. Besides, it achieved persistence by modifying the registry via the WinAPI function \texttt{RegSetValueExA()}. Later, the malicious DLL collected process information by calling the WinAPI functions \texttt{CreateToolhelp32Snapshot()} and \texttt{Process32First()}. Finally, it downloaded another malicious DLL to search for credentials in Outlook.

21 out of 22 EDR systems blocked this attack at different steps. Most blockages happen at the first step when the Explorer executes the Word document. This behavior is already pretty suspicious, as it downloaded an untrusted file. In the following steps, Word downloaded a malicious DLL file and a malicious VBS file and then executed it. The malicious intention is evident at this step, as this is a typical download and execution behavior. However, not all EDR systems responded to it: 19 EDR systems blocked the process, while three EDR systems either waited until later to block or didn't react. We checked their connectivity and found they could see the connected attack steps corresponding to this protection test. Although they have reasonable detection performance on single steps, these EDR systems failed to chain steps together to better understand the kill chain to provide appropriate protection.

% [0]:9
% [0,1,2]:4
% [0,1,2,3]:4
% [0,1]:2
% [0-6]:1
% [0-15]:1

\noindent\textbf{Scenario 2: TrickBot Execution, Discovery, and Kerberoasting.}

Fig.~\ref{wizard-spider-attack-2-2} in the Appendix shows a detailed attack graph of this scenario. In this scenario, the adversary authenticated into the victim's host using stolen credentials from scenario 1. Then, the adversary downloaded and executed a malicious EXE derived from TrickBot. The malicious EXE first established a C\&C session with an adversary-controlled server. Then, it collected various system information by executing shell commands. Finally, it downloaded a tool called \texttt{rubeus} to perform Kerberoasting~\cite{crowdstrike_kerberoasting}, which could steal encrypted credentials.

21 out of 22 EDR systems blocked this attack at different steps. In this scenario, the adversary connected to the target via RDP protocol as the first step. Looking at this step alone, it could be normal behavior. However, in the latter steps, the adversary downloaded a malicious file and executed it to establish a communication channel with the C\&C center. The malicious intention is evident at this step as it downloaded and executed an unknown file and established a suspicious outward connection. Only half of the EDR systems decided to block the process at steps 3 and 4. The rest of the EDR systems block the process when collecting system information in the later steps.

Although scenarios 1 and 2 had the same protection rate eventually, there is a noticeable delay in scenario 2 compared to scenario 1. One reason could be the difference in step visibility. As shown in Fig. \ref{wizard-spider-attack-2-1}(b) and \ref{wizard-spider-attack-2-2}(b), early steps in scenario 1 had better visibility than early steps in scenario 2. The third step in scenario 2 only had 64.29\% visibility, making it hard for EDR systems to gather enough information and react.

\begin{center}
\fcolorbox{black}{gray!10}{\parbox{.96\linewidth}{

\noindent\textbf{Finding 1:} Attack graph level correlation capabilities are necessary to achieve good defense because isolated single steps cannot provide enough confidence for EDR systems to respond. 

}}
\end{center}

% \noindent\textbf{Finding 1:} Attack graph level correlation capabilities are necessary to achieve good defense because isolated single steps don't provide enough confidence for EDR systems to react. 

The steps in Tests 3, 4, 5, and 6 happened as a connected kill chain on the same host in the detection test but are isolated into different test scenarios in the protection tests. This gives us a chance to investigate how isolated scenarios can affect defense. Since isolated scenarios contain fewer steps for correlating, the defense and response decisions primarily rely on single-step detection. We observed a significant drop in protection rate in tests 3, 4, and 5 as shown in Table \ref{protectiontesttable}. Test 4 only contained two steps that dumped system information (C disk and the registry) and received the lowest protection rate. Although dumping the entire C disk and the registry seems suspicious, such behaviors alone are usually not malicious enough to be escalated to alarms. Admittedly, we observed many cases in which EDR systems take action as soon as a suspicious file is downloaded and executed. Still, such a download and execution pattern wouldn't work well against file-less attacks, living-off-the-land attacks, and other evasion techniques.

Furthermore, some evidence shows EDR systems with poor performance didn't have graph-level correlation capabilities. For example, the detection screenshots from vendors like Deep Instinct didn't present any graph-level information along with the detection. In contrast, the detection screenshots from vendors like Sentinel One complement the detection with kill chain information on a graph.

\begin{center}
\fcolorbox{black}{gray!10}{\parbox{.96\linewidth}{

\noindent\textbf{Finding 2:} Although some EDR systems demonstrated good attack graph level correlation capabilities, we still identified three practical problems: delay in protection, lack of protection, and lack of cross-host correlation capability.

}}
\end{center}

Delay in protection problem exists ubiquitously in all scenarios. In the cases we analyzed, the adversary will log into the target and download a payload. More than half of the protection happens here, but the rest would happen either after the adversary had done some malicious behaviors or not at all. We checked their visibility. Most of them can see a connected attack chain. Those EDR systems require a longer kill chain to accumulate confidence before blocking a process. Such a mechanism prevents them from reacting quickly to threats. Sometimes, this even prevents them from reacting at all.

Lack of protection problems would still occur when some stealthy steps are applied. Besides requiring a longer chain to reach their confidence level, some EDR systems are susceptible to attack evasion techniques. For example, Test 3 modified the registry to achieve persistence. Tests 4 and 5 mimic system administrators to dump system information and modify system configurations. These tests applied more stealthy and sophisticated approaches than other tests, thus receiving a relatively low protection rate.

Attack graph level correlation should be applied on individual hosts and across hosts. In this evaluation, the adversary used the same tools and adopted similar attack patterns on different hosts in tests 2 and 3, respectively. Given test 2 happened before test 3, the protection performance in test 3 is not better than test 2. It suggests the EDR systems could not learn from the happened attacks to react to similar attacks in the future. Furthermore, no evidence shows that detection and response mechanisms use information across hosts to improve defensive performance.

\section{Overall Trend Analysis}
\label{sec:trend}
% \begin{itemize}
%     \item Participation in these evaluations has increased by nearly three times since the first round
%     \item No significant market-wide improvements were seen, both across all vendors and those who participated in all three rounds
%     \item Low-signature, stealthy techniques were the  hardest to detect and enrich in each round
% \end{itemize}

\begin{figure*}[!h]
\vspace{-1em}
     \centering
     \begin{subfigure}[b]{0.33\textwidth}
        \centering
        \includegraphics[width=\linewidth]{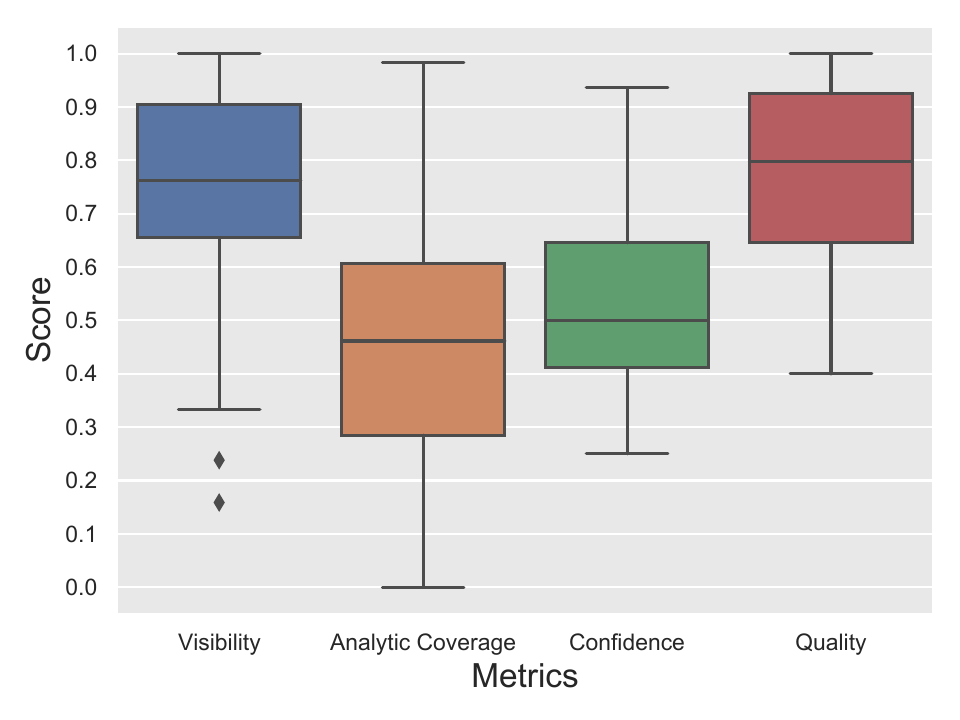}
        \caption{APT29 (2019)}
        \label{boxplot_apt29_tactic}
        \vspace{-1em}
     \end{subfigure}
     \hfill
     \begin{subfigure}[b]{0.33\textwidth}
        \centering
        \includegraphics[width=\linewidth]{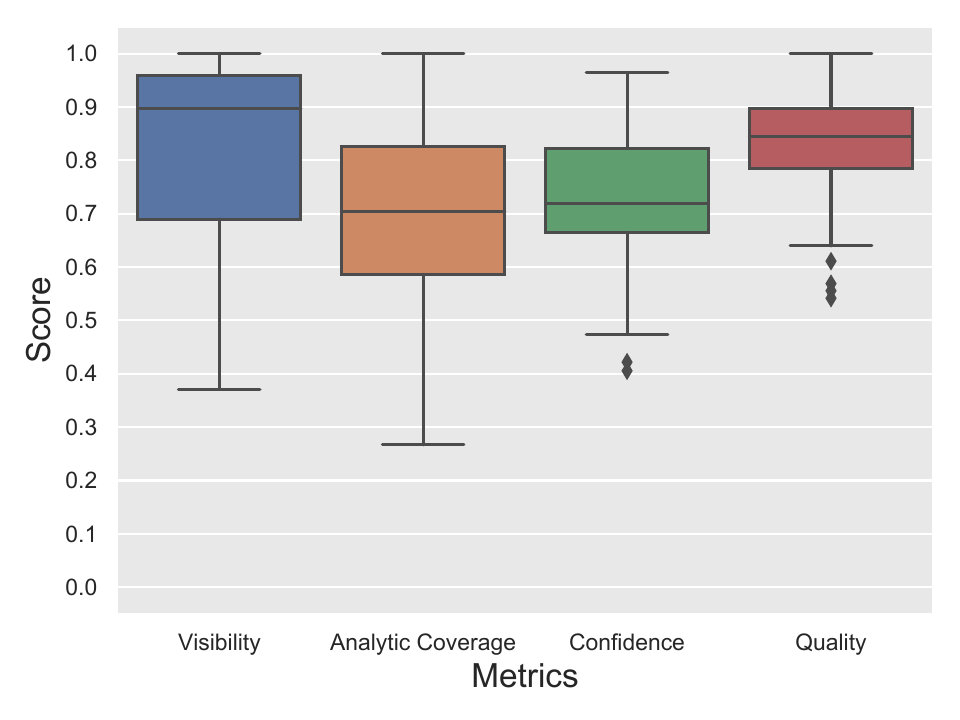}
        \caption{Carbanak+Fin7 (2020)}
        \label{boxplot_carbanak_fin7_tactic}
        \vspace{-1em}
     \end{subfigure}
     \hfill
     \begin{subfigure}[b]{0.33\textwidth}
        \centering
        \includegraphics[width=\linewidth]{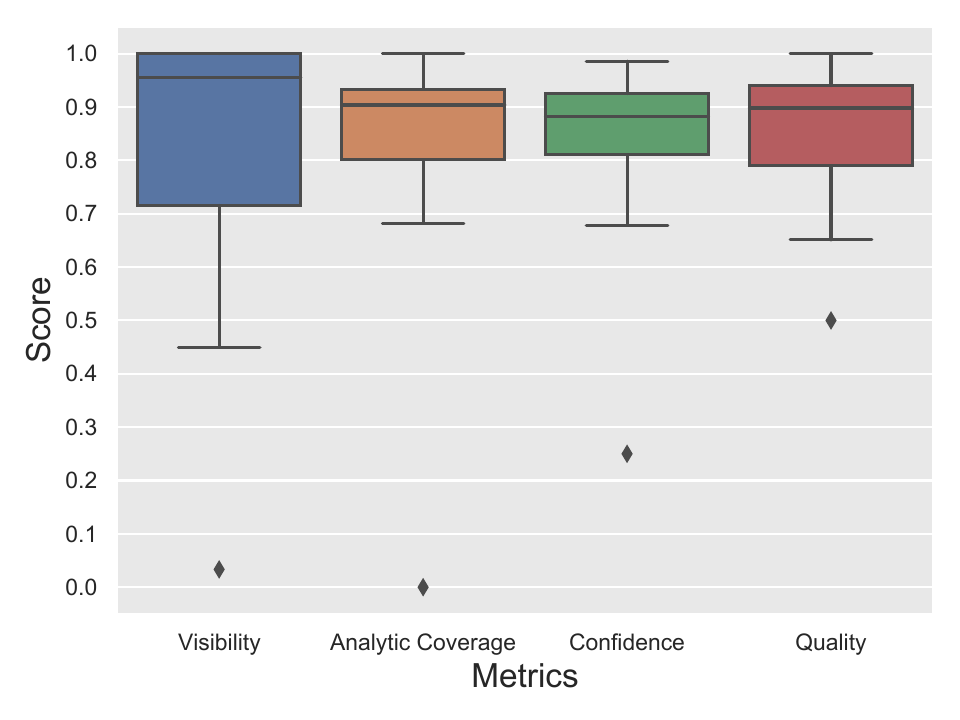}
        \caption{Wizard Spider+Sandworm (2022)}
        \label{boxplot_wizard_spider_sandworm_tactic}
        \vspace{-1em}
     \end{subfigure}
        \caption{Technique perspective score distribution of each metric in different evaluations. The metrics are visibility (blue), analytic coverage (orange), confidence (green), and quality (red) from left to right, respectively.}
        \label{fig:boxplot_tactic}
        \vspace{-1em}
\end{figure*}

\begin{figure*}[!h]
     \centering
     \begin{subfigure}[b]{0.33\textwidth}
        \centering
        \includegraphics[width=\linewidth]{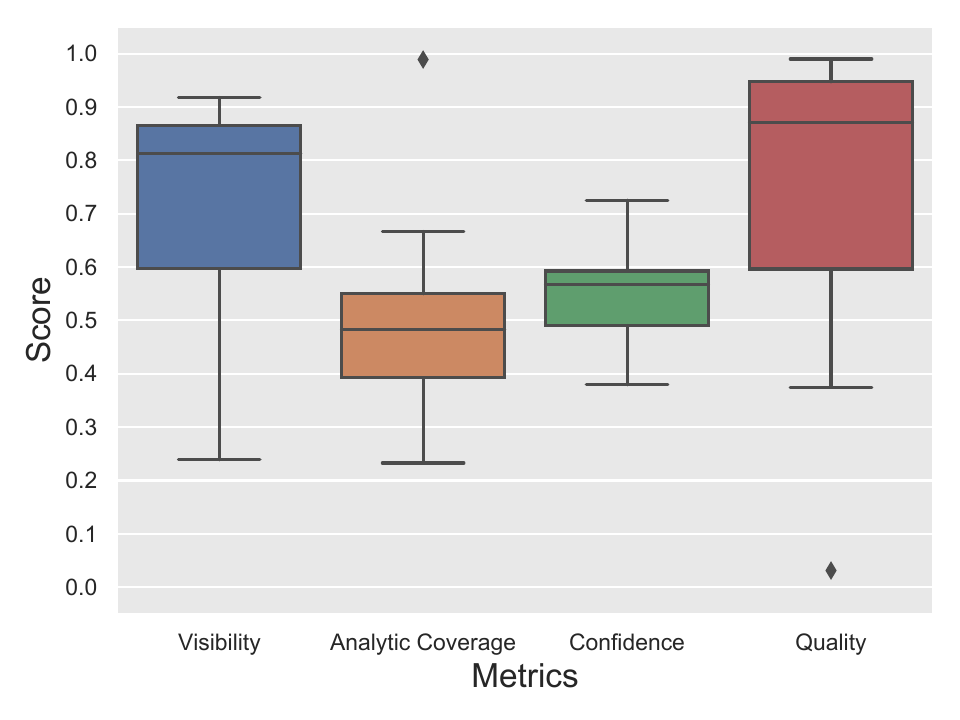}
        \caption{APT29 (2019)}
        \label{boxplot_apt29_vendor}
        \vspace{-1em}
     \end{subfigure}
     \hfill
     \begin{subfigure}[b]{0.33\textwidth}
        \centering
        \includegraphics[width=\linewidth]{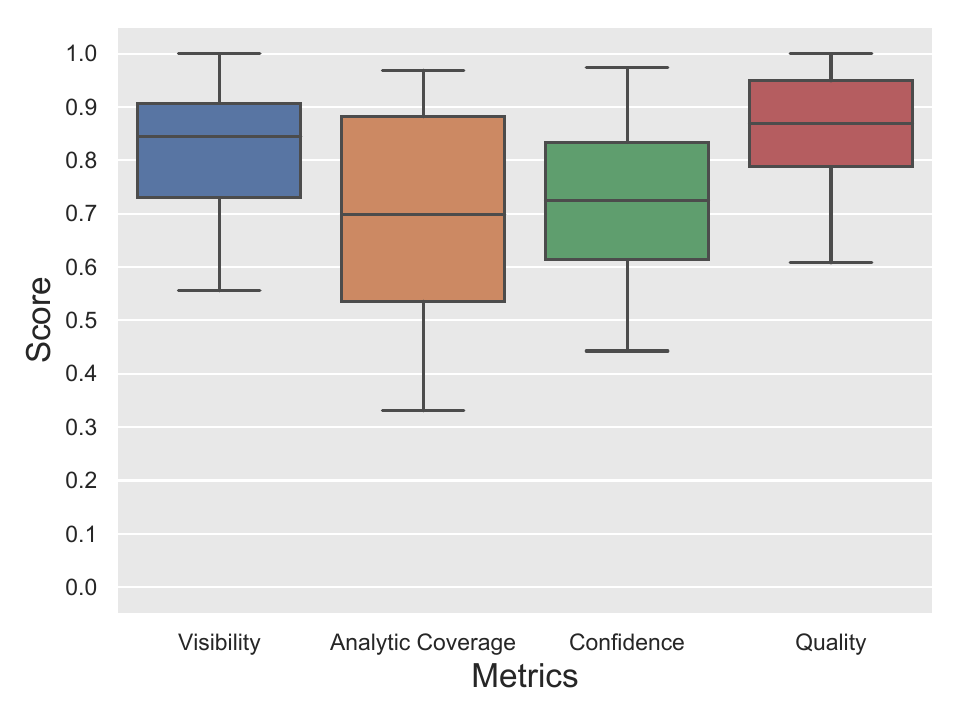}
        \caption{Carbanak+Fin7 (2020)}
        \label{boxplot_carbanak_fin7_vendor}
        \vspace{-1em}
     \end{subfigure}
     \hfill
     \begin{subfigure}[b]{0.33\textwidth}
        \centering
        \includegraphics[width=\linewidth]{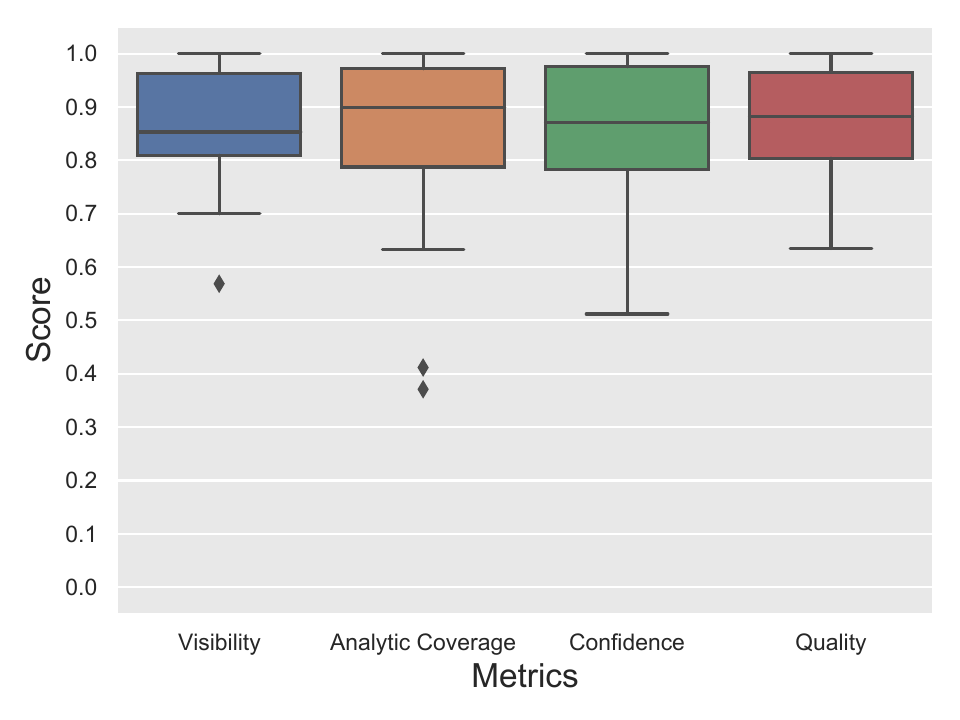}
        \caption{Wizard Spider+Sandworm (2022)}
        \label{boxplot_wizard_spider_sandworm_vendor}
        \vspace{-1em}
     \end{subfigure}
        \caption{Vendor perspective score distribution of each metric in different evaluations. The metrics are visibility (blue), analytic coverage (orange), confidence (green), and quality (red) from left to right, respectively.}
        \label{fig:boxplot_vendor}
        \vspace{-1em}
\end{figure*}

In this section, we examine several perspectives in all available datasets from MITRE Engenuity to investigate the paradigm of attacks and defenses in a real-world setting. We mainly analyze the results from more recent evaluations, especially the Carbanak+Fin7 (2020) and the Wizard Spider+Sandworm (2022) evaluations because they included APT emulations performed on multiple operating systems and they used a more established taxonomy to describe evaluation results compared to the previous evaluations. 

% \begin{figure}[!h]
% \hspace*{-.5cm}
% \centering
% \includegraphics[width=\linewidth]{figures/apt29/boxplot.pdf}
% \caption{Score distribution for each metric among all techniques in APT29 (2019) evaluation}
% \label{boxplot_apt29}
% \end{figure}

% \begin{figure}[!h]
% \hspace*{-.5cm}
% \centering
% \includegraphics[width=\linewidth]{figures/carbanak-fin7/boxplot.pdf}
% \caption{Score distribution for each metric among all techniques in Carbanak+Fin7 (2020) evaluation}
% \label{boxplot_carbanak_fin7}
% \end{figure}

% \begin{figure}[!h]
% \hspace*{-.5cm}
% \centering
% \includegraphics[width=\linewidth]{figures/wizard_spider-sandworm/boxplot.pdf}
% \caption{Score distribution for each metric among all techniques in Wizard Spider+Sandworm (2022) evaluation}
% \label{boxplot_wizard_spider_sandworm}
% \end{figure}

\subsection{Detection Coverage} 
\label{sec:detection_coverage}
% \textcolor{red}{Talk about MITRE's metrics using previous results.} 

We calculate the visibility and analytic coverage scores for individual techniques across the EDR systems participating in the evaluations. The distribution of visibility and analytic coverage scores from the technique perspective are shown in Fig.~\ref{fig:boxplot_tactic}.
% For instances, we show partial results of 15 top EDR systems and 16 well-covered techniques from the fine-grained analysis on the Wizard-Spider-Sandworm evaluation in Table \ref{partial-metrics-vendors} and \ref{partial-metrics-techniques}, respectively. Note that the Protection metric is not included in the table since not every EDR system participated in the protection test. We will discuss protection metric separately. 
We also calculate the visibility and analytic coverage scores for individual EDR systems as shown in Fig.~\ref{fig:boxplot_vendor}.
Then, we use these two metrics to analyze the overall trend of detection coverage. Moreover, we select a few EDR systems and techniques that receive excellent or very low scores and try to analyze the reasons behind them.

\subsubsection{Visibility}
As shown in Fig.~\ref{fig:boxplot_tactic} and \ref{fig:boxplot_vendor}, the distributions of technique visibility scores and vendor visibility scores are mostly skewed to the left. The median of technique visibility distribution from Wizard Spider+Sandworm (2022) evaluation is around 95\%, which means half of the attack steps can be seen by at least 95\% of the EDR systems in evaluations. The median vendor visibility distribution from the same evaluation is around 85\%, suggesting half of the EDR systems can see more than 85\% of the attack steps. Comparing the visibility score distribution in the three evaluations, we see an obvious improvement in the median and the lowest score for both techniques and vendors over the years.

Two techniques in Command and Control (C\&C) receive low scores across the EDR systems in the Carbanak+Fin7 evaluation. Specifically, only around 40\% of the EDR systems can record \textit{Encrypted Channel} or \textit{Application Layer Protocol} communications. In this evaluation, such communications include transmitting data over SSH protocol, MSSQL transactions, and HTTPS protocol. Meanwhile, the visibility of other C\&C techniques, which establish the connection via TCP, are all above 80\% among the EDR systems. Based on this observation, we conclude most EDR systems selectively collect network traffic data on the transport layer and ignore the application layer protocols.

Surprisingly, 15 techniques achieved perfect coverage among all EDR systems in the Wizard Spider+Sandworm evaluation, including six discovery techniques, three Defensive Evasion techniques, and a few techniques in other tactics. Those techniques involve manipulating or investigating system configurations and services, which implies all EDR systems have no trouble monitoring system configurations and services. In addition, four out of six techniques in the Execution Tactic receive a visibility score above 90\% in the Carbanak+Fin7 evaluation. Those techniques involve a process loading certain libraries or executing specific commands. We conclude that all EDR systems emphasize monitoring process loading and execution so that the techniques related to execution achieve the best visibility among all techniques.

SentinelOne achieves 100\% visibility on all techniques used in the Carbanak+Fin7 evaluation, while only around 50\% of the techniques are visible to AhnLab in the same evaluation. AhnLab hardly monitors the file system, as most of the techniques associated with \textit{Collection} and \textit{Credential Access} remain invisible to AhnLab. Furthermore, AhnLab does not monitor network activities except for a few file download behaviors and some TCP connections. It is particularly interesting that AhnLab can see and even raise alarms on some file download behaviors by Powershell but remain completely blind to the rest of the file download behaviors by Powershell from the same IP.

We observe changes and continuities by comparing the visibility score between the Carbanak+Fin7 and Wizard Spider+Sandworm evaluations. Most EDR systems obtain higher visibility scores in the latter evaluation, especially AhnLab, whose visibility score has a huge boost from 0.517 to 0.761. This implies the entire industry has improved in making various behaviors visible. The visibility difference in techniques is interesting. The two evaluations have about the same average visibility scores (about 80\%). However, some techniques like \textit{Archive Collected Data} have a perfect visibility score in the Carbanak+Fin7 evaluation but only receives a 0.033 visibility score in the Wizard Spider+Sandworm evaluation, which means it's only visible to one out of 30 EDR systems. The dramatic discrepancy suggests technique is not an appropriate unit for detection coverage since the same technique could be implemented with totally different procedures and consequently require distinct detection capabilities.

% To sum up, we have the following findings from analyzing visibility metrics.

\begin{center}
\fcolorbox{black}{gray!10}{\parbox{.96\linewidth}{

\noindent\textbf{Finding 3:} Most EDR systems have good data collection capability, and this capability is improving every year. In the most recent Wizard Spider+Sandworm (2022) evaluation,  75\% of the EDR systems can identify more than 80\% of the attack steps.

% Some techniques are easy to monitor and distinguish from benign behaviors, such as modifying system services and executing malicious software. Others like \textit{Encrypted Channel} and \textit{Archive Collected Data} are hard to discern at the system provenance level without additional context. Thus, many EDR systems choose to ignore such techniques to reduce false positive alarms.

}}
\end{center}

\begin{center}
\fcolorbox{black}{gray!10}{\parbox{.96\linewidth}{

\noindent\textbf{Finding 4:} Large discrepancies in visibility for the same technique in different evaluations suggest that techniques are still too coarse-grained for detection coverage. A more fine-grained unit, such as generalized technique implementations, is needed.

}}
\end{center}

\subsubsection{Analytic Coverage}
As shown in Fig.~\ref{fig:boxplot_tactic} and \ref{fig:boxplot_vendor}, analytic coverage has significantly improved over the years.  In the most recent Wizard Spider+Sandworm (2022) evaluation, 50\% of visible attack steps would trigger alarms on more than 90\% of the EDR systems, and 50\% of the EDR systems would generate alarms for at least 90\% of the visible attack steps. Unlike the distribution of visibility, analytic coverage exhibits a nearly symmetric distribution. 

The techniques receiving low analytic scores are mainly in C\&C, Exfiltration, and Collection tactics. Not surprisingly, \textit{Archive Collected Data} has an analytic coverage score of 0 given its low visibility score, which means only telemetry detection is made for this technique in the Wizard Spider+Sandworm evaluation. Other techniques like \textit{Exfiltration Over C\&C Channel} and several C\&C techniques also received below 50\% analytic coverage scores. Although such behaviors might be indistinguishable from everyday benign behaviors, they could still be identified as a part of a complete attack chain. Thus, detecting these behaviors challenges EDR systems' capabilities of assembling attack chains from scattered individual events. Low analytic coverage scores on those techniques suggest all EDR systems have room for improvement in their event correlation capability.

Moreover, the techniques receiving high analytic coverage scores mostly fall into categories of Defense Evasion and Credential Access, including \textit{OS Credential Dumping}, \textit{Inhibit System Recovery} and \textit{Process Injection}. Such techniques are easy to separate from normal behaviors as they carry malicious intentions conspicuously.

% Sophos and Cisco in the Carbanak+Fin7 evaluation and Rapid7 in the Wizard Spider+Sandworm evaluation have low analytic coverage scores (below 40\%) across all EDR systems, which is unsurprising as they also have relatively low visibility scores. By examining the detection they made beyond the telemetry level, we found they specialize in detecting a small set of techniques. Sophos has decent coverage on \textit{Disabling Firewall},\textit{Command Line Execution}, \textit{Registry Key Modification} and \textit{Screen Capture}; Cisco well covers \textit{Registry Key Modification}, \textit{OS Credential Dumping} and \textit{Ingress Tool Transfer}; Rapid7 does a good job covering system credential discovery like \textit{OS Credential Dumping}. These techniques all have high analytic coverage scores, which suggests they are relatively easy to detect due to their uniqueness. We conclude that EDR systems like Sophos and Cisco choose risk-averse strategies to limit false positive alarms. Although such strategies might be reasonable in some scenarios, they should challenge themselves to broaden the spectrum of alarm generation instead of being satisfied with merely collecting data.

Comparing the analytic coverage scores between the Carbanak+Fin7 and Wizard Spider+Sandworm evaluations, the analytic coverage scores have a significant improvement for all techniques as the average increases from 69.8\% to 85.5\% and the standard deviation decreases from 16\% to 8\%. Techniques in C\&C, Exfiltration, and Collection tactics that receive less than 60\% analytic coverage scores in the Carbanak+Fin7 evaluation mostly have higher than 70\% analytic coverage scores in the Wizard Spider+Sandworm evaluation. 

% To sum up, we have the following findings from analyzing analytics metrics.

\begin{center}
\fcolorbox{black}{gray!10}{\parbox{.96\linewidth}{

\noindent\textbf{Finding 5:} Although EDR systems' ability to determine malicious behaviors improves over time, they struggle to detect `living-off-the-land' threats. Throughout all evaluations, several techniques, including \textit{Encrypted Channel}, \textit{Exfiltration Over C\&C Channel}, and \textit{Archive Collected Data}, etc., consistently had worse detection rates. It is hard for EDR systems without event correlation capabilities to discern whether such individual steps are malicious at the system provenance level.

}}
\end{center}

Moreover, we divide EDR systems and techniques into four quadrants according to their visibility and analytic coverage scores for comparison. As shown in Fig.~\ref{detection_vs_visibility_vendors} and \ref{detection_vs_visibility_techniques} in the Appendix. 

For EDR systems, the top performers like Sentinel One and Palo Alto Networks at the top right corner are excellent at visibility and detection, and the bottom performers like AhnLab at the bottom left corner need to catch up with both capabilities. Interestingly, there are some ramifications among the EDR systems in the middle. The five EDR systems in the top left quadrant, like FireEye, tend to focus on detection capability. FireEye has way above average analytics score (90.3\% vs. 69.8\%) but slightly below average visibility score (78.2\% vs. 80.8\%). On the contrary, the six EDR systems in the bottom right quadrant, like CrowdStrike, tend to put more emphasis on visibility capability. CrowdStrike has a higher-than-average visibility score (87.4\% vs. 80.8\%) but a lower-than-average analytic coverage score (46.2\% vs. 69.8\%).

For techniques, 42 out of 63 falls into the top right quadrant and the bottom left quadrant, meaning they have comparable visibility and analytic coverage scores. For instance, \textit{Network Share Discovery} has an excellent visibility score (100\%) and analytic coverage score (90.1\%), while \textit{Exfiltration Over C\&C Channel} has a poor visibility score (43.7\%) and analytic coverage score (22.6\%). Six out of 63 techniques fall into the top left quadrant. Those techniques receive below-average visibility scores but above-average analytic coverage scores. It implies it is hard for EDR systems to collect related data, but they can be easily identified as malicious behaviors once visible. For example, \textit{Access Token Manipulation} receives only a 42.9\% visibility score but an 80.6\% analytic coverage score. Such techniques often involve some defense evasion intentions, or it might be expensive to collect related data, making them naturally stealthy. 15 out of 63 techniques fall into the bottom right quadrant. Those techniques receive above-average visibility scores but below-average analytic coverage scores. It suggests they are easily visible to EDR systems, but it is hard to associate them with malicious behaviors. For example, \textit{Email Collection} has a 95.2\% visibility score but only a 37.5\% analytic coverage score. Such behaviors are not stealthy but blend in with benign behaviors a normal user would perform.

To sum up, 18 out of 29 EDR systems have comparable visibility and analytic coverage scores. In comparison, five of the rest focus more on their detection capability, and six EDR systems lean towards improving their visibility. 42 out of 63 techniques have comparable visibility scores and analytic coverage scores. Among the remaining 21 techniques, six techniques are hardly visible but easy to detect once they are visible. The other 15 techniques are easily visible but hard to identify as malicious behaviors.

\subsection{Detection Confidence}
% \textcolor{red}{List out percentage of detection made in each category. Use old confidence metric results.}

The layout of confidence scores looks similar to analytic coverage scores since they reflect the soundness of detection results. However, we propose the confidence metric in addition to the analytic coverage metric because the confidence metric takes different detection levels above telemetry into account. As a weighted average, the confidence score indicates the overall detection level made by an EDR system or against a technique. Along the spectrum of confidence scores, a 25\% confidence score means the detection level is at telemetry on average; a 50\% confidence score means the detection is at General Behavior on average; a 75\% and 100\% confidence score means the detection level is at Tactic and Technique, respectively. For instance, \textit{Input Capture} has a perfect analytic coverage score, while it only has a 61.7\% confidence score in Carbanak+Fin7 (2020) evaluation, suggesting most of the detection levels are between General Behavior and Tactic. Other Credential Access techniques like \textit{Unsecured Credentials and Credentials from Password Stores} also have discrepancies in analytic coverage and confidence. 
% Fig.~\ref{confidence-analytics_techniques} in the Appendix shows the confidence-analytic difference across all techniques in the Carbanak+Fin7 evaluation.

Interestingly, the techniques that receive the highest and lowest confidence scores are all in Credential Access or Discovery tactics. \textit{OS Credential Dumping} and \textit{Network Share Discovery} receive above 90\% confidence scores, whereas \textit{Credentials from Password Stores} and \textit{Permission Group Discovery} obtain below 30\% confidence scores. All EDR systems can identify malicious behaviors related to \textit{OS Credential Dumping} and \textit{Network Share Discovery} and complement alarms with additional context like motivations and techniques used. On the contrary, EDR systems struggle to set off alarms for malicious behaviors related to \textit{Credentials from Password Stores} and \textit{Permission Group Discovery}, let alone providing additional context. 

% CheckPoint receive an astonishing 97.4\% confidence score while Sophos and Cisco receive confidence scores below 50\%. 
% Admittedly, the distribution of analytics and Confidence metrics are very similar, but there are some subtle discrepancies worth investigating. 
% We will analyze them later in Sec \ref{subsec:inter-metric}.

% As the weighted score of detection methods, confidence scores share some general trends as the analytic coverage scores with the existence of some notable discrepancies. Figure \ref{confidence-analytics_techniques} in the Appendix shows the confidence-analytic difference across all techniques. 

The technique \textit{Web Service} has a significantly lower confidence score than its analytic coverage score. This suggests the alarms on \textit{Web Service} fail to provide detailed contexts, like the motivation of such behavior or the specific technique used. Such vague alarms usually require system administrators to spend more time investigating and responding. On the other hand, techniques like \textit{Exfiltration Over C\&C Channel} remarkably higher confidence scores than its analytic coverage score, which implies detection on \textit{Exfiltration Over C\&C Channel} comes with a satisfying amount of details although the number of generated alarms is relatively low.

For EDR systems, the confidence scores and analytic coverage scores also share the general trend but with a few outliers. We calculate the confidence-analytic difference across all EDR systems participating in Carbanak+Fin7 (2020) evaluation to further investigate their detection confidence.
% Fig.~\ref{confidence-analytics_vendors} in the Appendix shows the confidence-analytic difference across all EDR systems that participated in Carbanak+Fin7 (2020) evaluation. 
CyCraft has a notably lower confidence score than the analytic coverage score, suggesting CyCraft puts more emphasis on detection coverage than other contexts. On the contrary, EDR systems like Sophos and AhnLab have much higher confidence scores than their analytic coverage score, which suggests they value the context provided in detection more than the coverage. Top performers like Palo Alto Networks, Sentinel One, and CheckPoint have about the same confidence score and analytic coverage score, which implies their detection has satisfying coverage and abundant details.

Comparing the distribution of confidence scores over the three years in Fig.~\ref{fig:boxplot_tactic} and \ref{fig:boxplot_vendor}, the technique median confidence score improved from 50\% in APT29 to 72\% in Carbanak+Fin7, and finally to 88.25\% in Wizard Spider+Sandworm. The confidence score distributions exhibited a decrease in range while shifting towards the higher end. This implies a significant enhancement in the level of detail with the detection. In the APT29 (2019) evaluation, only 50\% of the attack steps can be detected at the General Behavior level or above by all EDR systems on average; however, in the Wizard Spider+Sandworm (2022) evaluation, 75\% of the attack steps can be detected at the Technique level on average.

% To sum up, we have the following findings from analyzing confidence metrics.

\begin{center}
\fcolorbox{black}{gray!10}{\parbox{.96\linewidth}{

\noindent\textbf{Finding 6:} Different amounts of details in alarms reflect EDR systems' detection confidence. Throughout the five evaluations, EDR systems are less confident to trigger alarms on techniques that widely exist in everyday activities such as \textit{Email Collection}, \textit{Exfiltration Over C\&C Channel}, and \textit{Ingress Tool Transfer}. Thus, EDR systems tend to provide less contextual information, like their roles in the attack chain. On the contrary, EDR systems are more confident in detecting typical malicious behaviors like \textit{OS Credential Dumping} and \textit{Network Sniffing}. EDR systems' overall detection confidence has improved remarkably, as shown in Fig. \ref{fig:boxplot_tactic} and \ref{fig:boxplot_vendor}.
% EDR systems are more confident and provide more context for techniques like \textit{Exfiltration Over C\&C Channel} and \textit{Data from Local System}. However, little context can be provided for alarms on techniques like \textit{Web Service}. All in all, EDR systems' ability to complement alarms with comprehensive contextual information like motivations and techniques has been improving remarkably over the years.

}}
\end{center}

% \begin{itemize}
%     \item \textbf{EDR systems are more confident and provide more detection context for some techniques like \textit{Exfiltration Over C\&C Channel} and \textit{Data from Local System}; whereas little contextual is provided with detection on techniques like \textit{Web Service}.}
%     \item \textbf{EDR systems like CyCraft put emphasis on detection coverage to provide expansive coverage without detailed context; on the other hand, EDR systems like Sophos and AhnLab focus on detecting a small set of threats confidently with details. Top performers like PaloAltoNetworks, SentinelOne and CheckPoint have both satisfying coverage and a decent amount of details.}
% \end{itemize}

% \begin{itemize}
    % \item EDR systems fail to complement alarms with comprehensive context like motivations and techniques used. Again, EDR systems need better alert correlation capability to deal with such threats.
    % \item Although EDR systems can better detect malicious behaviors, their capabilities of providing corresponding context need to be further improved.
% \end{itemize}

% \subsection{Kill Chain Composition} 
% \textcolor{red}{List of frequent structure of kill chain}

\begin{table*}[]
\vspace{-0.5em}
\caption{Data Sources in MITRE Evaluations}
\vspace{-0.5em}
\label{tab:datasource}
\begin{tabular}{|l|c|c|}
\hline
Campaign                 & \# of Data Source & Top 5 Data Sources                                                                   \\ \hline
APT29 (2019)                   & 9                  & File, Command, Process, Script, Network Traffic                                  \\ \hline
Carbanak+FIN7 (2020)         & 25                 & Process, File, Network Traffic, Script, OS API Execution \\ \hline
Wizard Spider+Sandworm (2022) & 41                 & Process, File, Network Traffic, OS API Execution, Logon Session       \\ \hline
\end{tabular}
\vspace{-0.5em}
\end{table*}

\subsection{Detection Quality}
% \textcolor{red}{Investigate what techniques are often labeled with modifier. Use old quality metric results.}
We calculated the quality metric for techniques and EDR systems in the recent three evaluations, as shown in Fig.~\ref{fig:boxplot_tactic} and \ref{fig:boxplot_vendor}, respectively. The detection quality score has been increasing over the years. 
\textit{Credentials from Password Stores} receive the lowest quality score (54.2\%) in the Carbanak+Fin7 evaluation, which suggests significant delay and manual efforts are involved. In this scenario, the credentials stored in the Chrome web browser are accessed via a malicious tool. This technique has a fairly low Visibility score and analytic coverage score. Even when it is detected, it usually requires human analysis. Other techniques with low quality scores (below 60\%) include \textit{Inter-Process Communication}, \textit{Credentials from Password Store} and \textit{Data from Local System}, which require modifying detection policies depending on the local environments. We suspect failures in detecting this technique are also related to systems' weak ability to link individual events to an attack chain. While accessing the credentials stored in the browser itself doesn't seem very suspicious, downloading an unknown tool and using it to access credentials makes it very suspicious.

OpenText and DeepInstinct receive the lowest quality score (around 60\%) among the EDR systems, while some EDR systems like SentinelOne, ReaQta, and CyCraft obtain close perfect quality scores in the Carbanak+Fin7 evaluation. The score differences imply EDR systems' various self-adapting abilities. Systems with high quality scores can work effectively in different environments without much human intervention, while systems with low quality scores require a lot of manual tuning and analysis.

% \textcolor{red}{take a look at individual modifiers separately}

% To sum up, we have the following findings from analyzing quality metrics.

\begin{center}
\fcolorbox{black}{gray!10}{\parbox{.96\linewidth}{

\noindent\textbf{Finding 7:} EDR systems often require extra manual effort to detect techniques that are closely integrated with local environments, such as \textit{Credentials from Password Stores} and \textit{Inter-Process Communication}. 
% EDR systems have a wide range of self-adapting abilities to the environments they are deployed in.
Only four tested systems in the Wizard Spider+Sandworm (2022) evaluation do not require extra effort to detect such techniques.

}}
\end{center}

% \begin{center}
% \fcolorbox{black}{gray!10}{\parbox{.96\linewidth}{

% \noindent\textbf{Finding 8:} \textcolor{red}{talk about delay findings}

% }}
% \end{center}

% \begin{itemize}
%     \item EDR systems are not good at automatically detecting some techniques that require different detection configurations depending on the local environments, such as \textit{Credentials from Password Stores}, \textit{Inter-Process Communication} and \textit{Data from Local System}.
%     \item EDR systems have a wide range of self-adapting abilities to the environments they are deployed in. Only 3 or 4 EDR systems do not require any extra effort in each emulation, whereas other EDR systems need additional processing or manual analysis, more or less, to achieve their desired performance.
% \end{itemize}

\subsection{Data Source}
% \textcolor{red}{List out the frequency of data sources and their correlation with detection.}

The number of distinct data sources has been changing over the years. No data source information is available in the APT3 evaluation (2018). Since the APT29 evaluation (2019), MITRE has started to collect data source information in the detection results. As shown in Table \ref{tab:datasource}, nine distinct data sources are recorded in the APT29 evaluation results, whereas in the most recent Wizard Spider+Sandworm evaluation (2022), 41 different data sources are recorded. As the number of distinct data sources increases over the years, not only do the existing data sources become more specific, but some new data sources are also included in the data sources. At the same time, the taxonomy of data sources has been changing. In the APT29 and Wizard Spider+Sandworm evaluations, the data sources are recorded in \textit{category: sub-category} format.
In contrast, in the Carbanak+FIN7 evaluation, the data sources are recorded as \textit{category} without further sub-categories. In the APT29 evaluation, the data sources are from the process, file, registry key, and network connection creations, as well as script and command line executions. In the Wizard Spider+Sandworm evaluation, network-related data sources include network connection creation, traffic content, and traffic flow. Besides, additional data sources, such as firewall metadata and network share access, are included. Our findings in \S\ref{sec:detection_coverage} demonstrated such enrichment in data sources has a positive correlation with the improvement in the detection performance over the years.
% Table xx shows the frequency of data sources appeared in all detection results.

% We have the following findings from investigating the evolution of data sources used in evaluations.

\begin{center}
\fcolorbox{black}{gray!10}{\parbox{.96\linewidth}{

\noindent\textbf{Finding 8:} Increasing complexity and variety of data sources suggest EDR systems can utilize extensive information from different dimensions. Although an increasing number of data sources are used, the top data sources remain unchanged. Process, file, network, scripts, and system calls/APIs are still the most fundamental and valuable data sources for EDR systems.

}}
\end{center}

\subsection{Compatibility}
% \textcolor{red}{Observe the platform compatibility in the last 2 evaluations as they used Windows and Linux platforms. Look at protection support.}
In the first two evaluations, APT3 and APT29, the target environments only contain Windows hosts; thus, all participants must support the Windows platform.
Starting from the Carbanak+Fin7 evaluation in 2020, MITRE enrich the variety of target environments by including Linux servers as parts of the target system.
In the Carbanak+Fin7 evaluation, 22 out of 29 participants supported the Linux platform. In the following Wizard Spider+Sandworm evaluation, 22 out of 30 participants supported the Linux platform. 

In \S\ref{subsec:effectiveness}, we found EDR systems had a low protection rate against attacks on Linux. The attack on the Linux platform follows a similar pattern to the ones on Windows: uploading a payload and using it to establish C\&C connections. Given the attacks have similar visibility and attack pattern, most vendors can protect against the attacks on Windows, but the attacks on Linux were only blocked by around half of the EDR systems.

% \textcolor{red}{compare the overall detection performance in linux and windows}
% Improve data analysis and protection capability on the Linux platform. The attack on Linux platform follows a similar pattern with the ones on Windows: uploading a payload and using it to establish C\&C connections. However, having the similar visibility and similar attack pattern, the attacks on Windows can be protected by most of the vendors but the attacks on Linux were only blocked by around half of the vendors.

% In summary, we have the following findings after investigating the compatibility of EDR systems.

\begin{center}
\fcolorbox{black}{gray!10}{\parbox{.96\linewidth}{

\noindent\textbf{Finding 9:} Data collection and protection capability needs improvement on the Linux platform since around 25\% of the evaluated EDR products don't support the Linux platform. For similar attack patterns, EDR products present worse protection results on Linux than on Windows.

}}
\end{center}

\section{Related Work}
\label{sec:related}
% Much effort has been made into security benchmarks.

\subsection{Endpoint Detection and Response (EDR)} 
An increasing number of researches on endpoint security solutions have been conducted to improve APT defense methods and forensics technologies on various platforms. 
% TODO: list EDR work on different platforms: OS, network, cloud, mobile
EDR frameworks like HOLMES~\cite{HOLMES}, Poirot~\cite{poirot}, MORSE\cite{morse}, and others~\cite{Conan, sleuth, Unicorn, NoDoze, ProTracer, Wang2020, pagoda, pgaussian, zeng2021watson, zengy2022shadewatcher, alsaheel2021atlas} aim to improve defense on Windows and Linux operating system, while other methods like RiskRanker~\cite{riskranker} and E-EMD~\cite{E-EMD} target security on mobile and cloud platforms, respectively. However, they all use statistical measurements like false positive and true positive, precision, recall, accuracy, and F-Score, to describe the detection performance. They also include CPU and memory usage as measurements for overhead. However, it is hard to compare the measurements from different works due to the different datasets and hardware used to carry out the measurements. Surveys on the EDR systems~\cite{zipperle, li2021threat, apt_survey} mainly focus on the methodology and dataset used but pay little attention to EDR evaluation.

In efforts to improve security evaluations, researchers have been studying the evaluation flaws in existing security works. For instance, Van Der Kouwe et al.\cite{SoK} identifies a list of common benchmarking mistakes and indicates that benchmarking flaws exist widely in system security papers published in top venues, which suggests the necessity of standardizing security benchmarks. Following those papers' insights and suggestions, we propose our evaluation and interpretation framework for system security work in academia.

\subsection{Security Benchmark} 
Most existing works mainly focus on generating representative data sets since quality security data sets are scarce. As early as 1998, DARPA launched its intrusion detection evaluation~\cite{IDS} in collaboration with Lincoln Lab at MIT. Zuech et al.~\cite{NIDS_dataset} tried to generate network-based data sets for evaluating network intrusion detection systems (NIDS); Divekar et al.~\cite{ANIDS_dataset} modified existing network-based data sets to improve training performance in anomaly-based NIDS; Almakhdhub et al.~\cite{IoT_benchmark} targets benchmarking Internet of Things (IoT) devices. Additionally, the data set from the DARPA Transparent Computing program~\cite{darpa_tc} has been used widely in recent security work. Still, data from only two out of five attack campaigns are publicly available, and thus the attack scenarios are minimal. Although such data sets help mitigate the deficit in security evaluation data, they do not provide extensive methodologies for interpreting the results. 

Some other work aims to improve the explainability of evaluations. Hao et al.~\cite{SAST_benchmark} and Mendes et al.~\cite{web_system_benchmark} designed methodologies to obtain more explainable evaluation results for static application security testing (SAST) tools and web serving systems, respectively. However, their methods are specific to the targeting tools or systems. Thus, it is hard to expand the methodologies to other security fields.

Recently, some security studies have used the knowledge base built by MITRE ATT\&CK. Choi et al.~\cite{attack_sequence} used the tactic, technique, and procedure (TTP) proposed by MITRE ATT\&CK to generate attack sequences. On the other hand, Outkin et al.~\cite{defender} use attacks emulated in MITRE ATT\&CK evaluation as the attack models to discuss defender policy and resource allocation. Although they used MITRE ATT\&CK knowledge base, they didn't analyze and interpret MITRE ATT\&CK evaluations.

Other commercial security `benchmarks' apply miscellaneous self-designed metrics in various testing environments, purely focusing on comparing EDR vendors on behalf of the customers for marketing purposes. For instance, Gartner tries to address the benchmarking challenges with its Magic Quadrant~\cite{noauthor_magic_nodate}. However, the methodology is not transparent and is lack of explanation. Moreover, Gartner emphasizes secondary concerns for businesses like value and viability, which don't provide insights on improving the security systems' performance. Another attempt in the industry is AV-Comparatives~\cite{noauthor_avcomparatives_nodate}, which evaluates the anti-virus capabilities of security products. However, the evaluation methodology is not transparent like Magic Quadrant's, and the evaluations only focus on a narrow range of attack techniques. 

\section{Discussion}
\label{sec:discussion}
% \subsection{Other Problems in MITRE Evaluation}
    
% Admittedly, using step-wise detection coverage is an effective approach to standardize measurement results across platforms. However, missing further context makes it hard to perform an in-depth analysis. 
An important problem we could not address in this paper is missing information, including but not limited to false positive alarm volume, response time, and raw data. 

False positive alarm volume is an important indicator of manpower needed for using the EDR system~\cite{99FP, fuzzy_fatigue}. Low false positive volume means most alarms are true positive so that the system administrator can focus on mitigation. On the other hand, a high false alarm volume means many of the alarms are false positive alarms. Hence, the system administrator must discover true positive alarms before mitigating attacks, often leading to a needle-in-a-haystack problem.
Response time indicates the time elapsed between compromises and alarm generation, which measures the real-time capabilities of the EDR systems. Low response time means the system can detect threats fast so that the system administrator can keep the loss to a minimum. Although the delayed modifier gives some information about the delayed alarm, it does not provide quantitative data on how long the delay is.
% Containment capabilities measure how effectively the system stops threats without human intervention. Although in the latest MITRE evaluation release, protection tests have been added to account for containment capabilities, it only focuses on how EDR systems identify each threat, rather than act upon them. Killing the process might not always be the best action to take, especially when false positive alarms are triggered on essential processes. An EDR system with good containment capabilities should stop most threats effectively without hurting the normal operation of the system.
Moreover, MITRE only provides the detection results from the EDR systems, not the raw data such as system logs and network events. Missing the raw data prevents new EDR systems from using the same dataset to compare performance with existing ones and limits the information available to researchers when they analyze threats with poor detection coverage.
Missing the information described above hinders the analysis of EDR systems from many meaningful perspectives. We hope MITRE could include them in the future release of evaluations.

Another concern is the prospective compatibility of our interpretation framework within the context of the MITRE evaluation. 
While MITRE implemented notable modifications to its evaluation framework during the initial three rounds, the framework employed in the latest three rounds has demonstrated sustained consistency.
Our interpretation comprehensively encompasses all elements that have maintained this consistency throughout these rounds. Therefore, we are confident that our interpretation framework will remain pertinent and enduring in the foreseeable future.

\section{Conclusion}
\label{sec:conclusion}
By leveraging MITRE's evaluation efforts and introducing our analysis method, we offer valuable insights into the current capabilities of industrial EDR systems to bridge the gap between MITRE's raw evaluation results and comprehensive interpretations. This research aids researchers, practitioners, and vendors in understanding the strengths, limitations, and areas for improvement of EDR systems, ultimately enhancing enterprise security.

\begin{acks}
This material is based upon work supported by the National Science Foundation under grant no. 2148177 and is supported in part by funds from federal agency and industry partners as specified in the Resilient \& Intelligent NextG Systems (RINGS) program.
\end{acks}

% \nocite{*}
% \begingroup
% \raggedright
% \bibliographystyle{unsrt}
\bibliographystyle{ACM-Reference-Format}
\bibliography{citations}

% \endgroup
\appendix
\section{Appendix}
\label{sec:appendix}
\subsection{Ethics}
This study does not raise any ethical issues. All the datasets we used are publicly available and anonymized.

\subsection{Additional graphs}
\begin{figure}
\centering
\includegraphics[width=\linewidth]{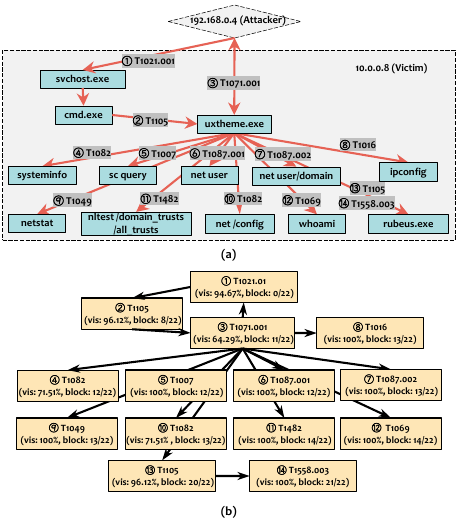}
\caption{The actual attack graph and the causal relationship attack graph for scenario 2 in Wizard Spider+Sandworm (2022) evaluation.}
\label{wizard-spider-attack-2-2}
\end{figure}

\begin{figure*}[]
\centering
\includegraphics[width=0.8\linewidth]{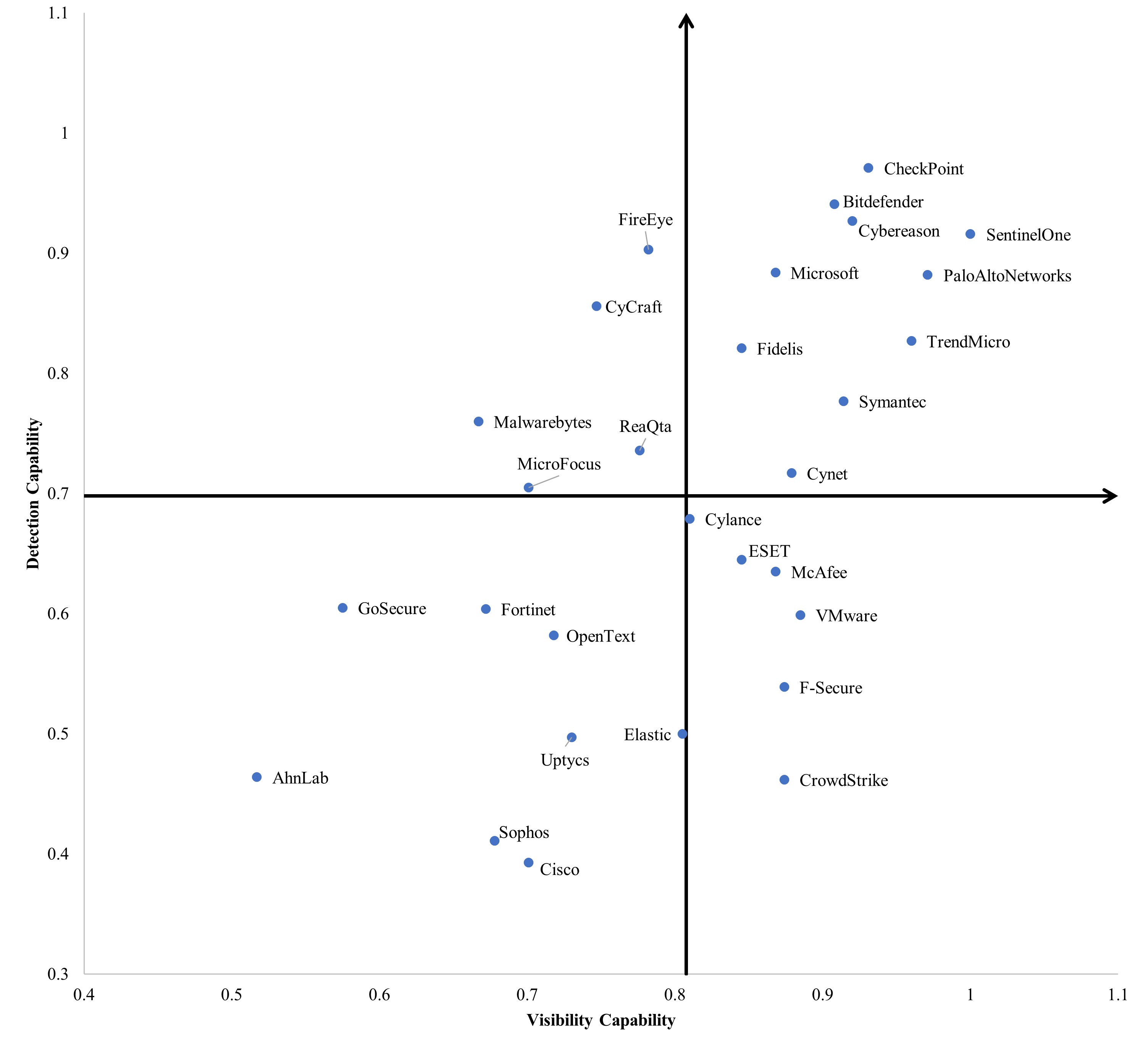}  
\caption{Analytics vs. Visibility Quadrant for EDR Systems}
\label{detection_vs_visibility_vendors}
\end{figure*}

\begin{figure*}[]
\centering
\includegraphics[width=0.9\linewidth]{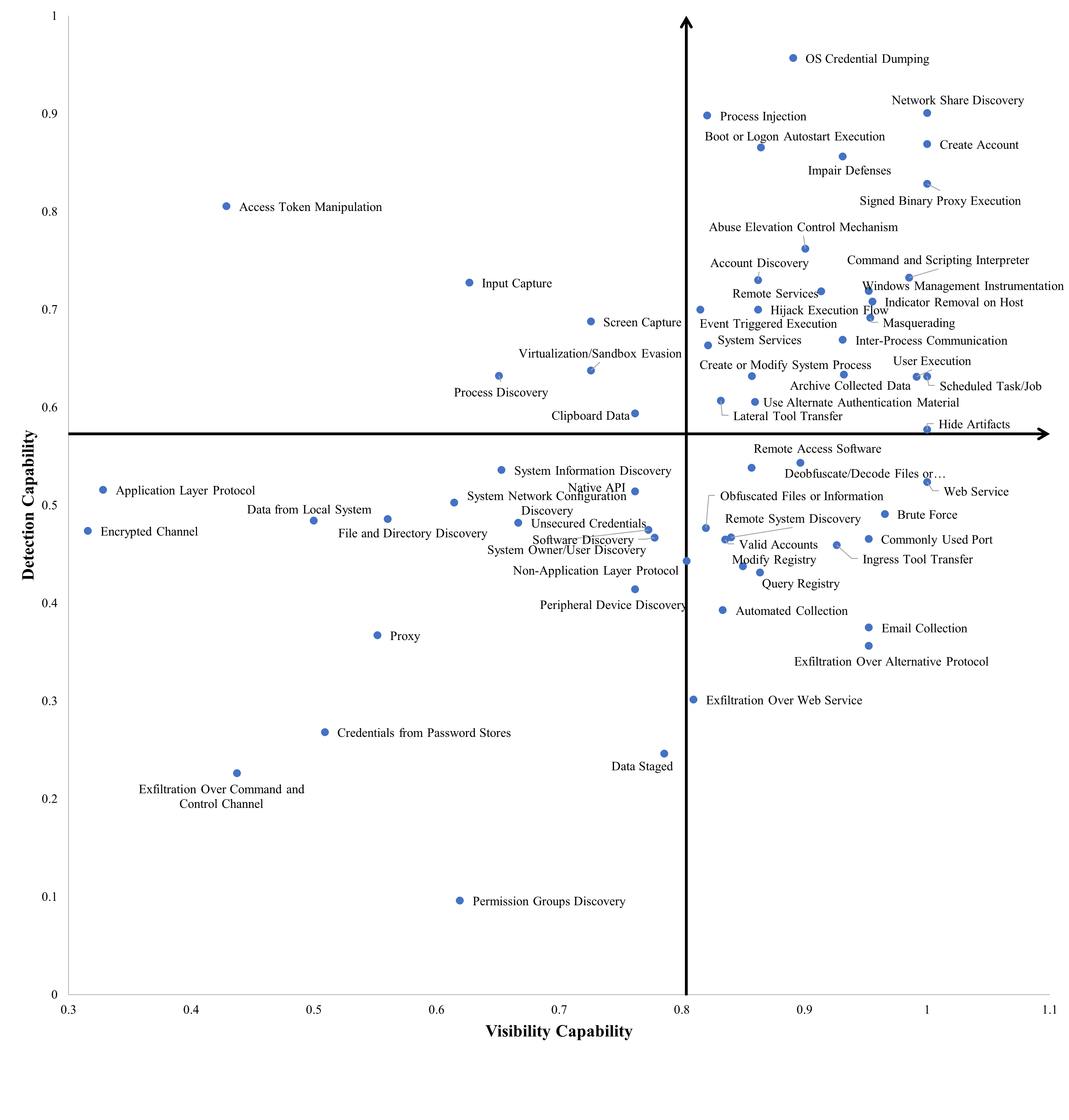}  
\caption{Analytics vs. Visibility Quadrant for Techniques}
\label{detection_vs_visibility_techniques}
\end{figure*}

Fig.~\ref{wizard-spider-attack-2-2} is another causal relationship attack graph we constructed. 
Fig.~\ref{detection_vs_visibility_vendors} shows the distribution of visibility and analytic scores of all EDR systems. Fig.~\ref{detection_vs_visibility_techniques} shows the same distribution of all techniques.

% \begin{figure*}[]
% \centering
% \includegraphics[width=0.8\linewidth]{figures/carbanak-fin7/vendor/confidence-analytics_vendors.png}  
% \caption{Confidence-Analytics Difference for EDR Systems}
% \label{confidence-analytics_vendors}
% \end{figure*}

% \begin{figure*}[]
% \centering
% \includegraphics[width=0.6\linewidth]{figures/carbanak-fin7/technique/confidence-analytics_techniques.png}  
% \caption{Confidence-Analytics Difference for Techniques}
% \label{confidence-analytics_techniques}
% \end{figure*}

\end{document}